\documentclass[longauth]{aa}

\usepackage{graphicx}
\usepackage{natbib}
\usepackage{scalerel}
\usepackage{adjustbox}
\usepackage[table]{xcolor}
\usepackage[caption=false]{subfig}
\usepackage{ulem}
\usepackage{placeins} 

\usepackage{txfonts}
\usepackage[pdfencoding=auto,psdextra]{hyperref}
\hypersetup{
    colorlinks=true,
    linkcolor=blue,
    filecolor=magenta,      
    urlcolor=blue,
    citecolor=blue
}
\makeatletter
\renewcommand*\aa@pageof{, page \thepage{} of \pageref*{LastPage}}
\makeatother

\usepackage[utf8]{inputenc}

\usepackage[switch, modulo]{lineno}

\usepackage{euclid}
\usepackage[nameinlink,capitalize]{cleveref}

\crefname{figure}{Fig.}{Figs.}
\Crefname{figure}{Figure}{Figures}
\crefname{section}{Sect.}{Sects.}
\Crefname{section}{Section}{Sections}

\renewcommand{\bea}{\begin{eqnarray}}
\renewcommand{\eea}{\end{eqnarray}}

\newcommand{\De}{\Delta}
\newcommand{\HH}{\mathcal{H}}

\newcommand{\de}{\mathrm{d}}

\newcommand\snr{\ensuremath{\text{S} / \text{N}}}

\NewDocumentCommand{\evalat}{sO{\big}mm}{%
  \IfBooleanTF{#1}
   {\mleft. #3 \mright|_{#4}}
   {#3#2|_{#4}}%
}

\newcommand{\zmean}{\bar{z}}

\begin{document}
%
%


\title{\Euclid: Relativistic effects in the dipole of the two-point correlation function\thanks{This paper is published on behalf of the Euclid Consortium.}}    


\newcommand{\orcid}[1]{} 
\author{F.~Lepori\thanks{\email{francesca.lepori2@uzh.ch}}\inst{\ref{aff1}}
\and S.~Schulz\orcid{0000-0002-8235-9986}\inst{\ref{aff1}}
\and I.~Tutusaus\orcid{0000-0002-3199-0399}\inst{\ref{aff2}}
\and M.-A.~Breton\inst{\ref{aff3},\ref{aff4},\ref{aff5}}
\and S.~Saga\orcid{0000-0002-7387-7570}\inst{\ref{aff6},\ref{aff7}}
\and C.~Viglione\inst{\ref{aff8},\ref{aff3}}
\and J.~Adamek\orcid{0000-0002-0723-6740}\inst{\ref{aff1}}
\and C.~Bonvin\orcid{0000-0002-5318-4064}\inst{\ref{aff9}}
\and L.~Dam\orcid{0000-0003-3163-9571}\inst{\ref{aff9}}
\and P.~Fosalba\orcid{0000-0002-1510-5214}\inst{\ref{aff8},\ref{aff3}}
\and L.~Amendola\orcid{0000-0002-0835-233X}\inst{\ref{aff10}}
\and S.~Andreon\orcid{0000-0002-2041-8784}\inst{\ref{aff11}}
\and C.~Baccigalupi\orcid{0000-0002-8211-1630}\inst{\ref{aff12},\ref{aff13},\ref{aff14},\ref{aff15}}
\and M.~Baldi\orcid{0000-0003-4145-1943}\inst{\ref{aff16},\ref{aff17},\ref{aff18}}
\and S.~Bardelli\orcid{0000-0002-8900-0298}\inst{\ref{aff17}}
\and D.~Bonino\orcid{0000-0002-3336-9977}\inst{\ref{aff19}}
\and E.~Branchini\orcid{0000-0002-0808-6908}\inst{\ref{aff20},\ref{aff21},\ref{aff11}}
\and M.~Brescia\orcid{0000-0001-9506-5680}\inst{\ref{aff22},\ref{aff23},\ref{aff24}}
\and J.~Brinchmann\orcid{0000-0003-4359-8797}\inst{\ref{aff25},\ref{aff26}}
\and A.~Caillat\inst{\ref{aff27}}
\and S.~Camera\orcid{0000-0003-3399-3574}\inst{\ref{aff28},\ref{aff29},\ref{aff19}}
\and V.~Capobianco\orcid{0000-0002-3309-7692}\inst{\ref{aff19}}
\and C.~Carbone\orcid{0000-0003-0125-3563}\inst{\ref{aff30}}
\and J.~Carretero\orcid{0000-0002-3130-0204}\inst{\ref{aff31},\ref{aff32}}
\and S.~Casas\orcid{0000-0002-4751-5138}\inst{\ref{aff33},\ref{aff34}}
\and M.~Castellano\orcid{0000-0001-9875-8263}\inst{\ref{aff35}}
\and G.~Castignani\orcid{0000-0001-6831-0687}\inst{\ref{aff17}}
\and S.~Cavuoti\orcid{0000-0002-3787-4196}\inst{\ref{aff23},\ref{aff24}}
\and A.~Cimatti\inst{\ref{aff36}}
\and C.~Colodro-Conde\inst{\ref{aff37}}
\and G.~Congedo\orcid{0000-0003-2508-0046}\inst{\ref{aff38}}
\and L.~Conversi\orcid{0000-0002-6710-8476}\inst{\ref{aff39},\ref{aff40}}
\and Y.~Copin\orcid{0000-0002-5317-7518}\inst{\ref{aff41}}
\and F.~Courbin\orcid{0000-0003-0758-6510}\inst{\ref{aff42},\ref{aff43}}
\and H.~M.~Courtois\orcid{0000-0003-0509-1776}\inst{\ref{aff44}}
\and H.~Degaudenzi\orcid{0000-0002-5887-6799}\inst{\ref{aff45}}
\and G.~De~Lucia\orcid{0000-0002-6220-9104}\inst{\ref{aff13}}
\and F.~Dubath\orcid{0000-0002-6533-2810}\inst{\ref{aff45}}
\and C.~A.~J.~Duncan\orcid{0009-0003-3573-0791}\inst{\ref{aff46}}
\and X.~Dupac\inst{\ref{aff40}}
\and S.~Dusini\orcid{0000-0002-1128-0664}\inst{\ref{aff47}}
\and M.~Farina\orcid{0000-0002-3089-7846}\inst{\ref{aff48}}
\and S.~Farrens\orcid{0000-0002-9594-9387}\inst{\ref{aff49}}
\and S.~Ferriol\inst{\ref{aff41}}
\and M.~Frailis\orcid{0000-0002-7400-2135}\inst{\ref{aff13}}
\and E.~Franceschi\orcid{0000-0002-0585-6591}\inst{\ref{aff17}}
\and S.~Galeotta\orcid{0000-0002-3748-5115}\inst{\ref{aff13}}
\and B.~Gillis\orcid{0000-0002-4478-1270}\inst{\ref{aff38}}
\and C.~Giocoli\orcid{0000-0002-9590-7961}\inst{\ref{aff17},\ref{aff18}}
\and A.~Grazian\orcid{0000-0002-5688-0663}\inst{\ref{aff50}}
\and F.~Grupp\inst{\ref{aff51},\ref{aff52}}
\and S.~V.~H.~Haugan\orcid{0000-0001-9648-7260}\inst{\ref{aff53}}
\and W.~Holmes\inst{\ref{aff54}}
\and F.~Hormuth\inst{\ref{aff55}}
\and A.~Hornstrup\orcid{0000-0002-3363-0936}\inst{\ref{aff56},\ref{aff57}}
\and S.~Ili\'c\orcid{0000-0003-4285-9086}\inst{\ref{aff58},\ref{aff2}}
\and K.~Jahnke\orcid{0000-0003-3804-2137}\inst{\ref{aff59}}
\and M.~Jhabvala\inst{\ref{aff60}}
\and E.~Keih\"anen\orcid{0000-0003-1804-7715}\inst{\ref{aff61}}
\and A.~Kiessling\orcid{0000-0002-2590-1273}\inst{\ref{aff54}}
\and M.~Kilbinger\orcid{0000-0001-9513-7138}\inst{\ref{aff49}}
\and B.~Kubik\orcid{0009-0006-5823-4880}\inst{\ref{aff41}}
\and M.~Kunz\orcid{0000-0002-3052-7394}\inst{\ref{aff9}}
\and H.~Kurki-Suonio\orcid{0000-0002-4618-3063}\inst{\ref{aff62},\ref{aff63}}
\and S.~Ligori\orcid{0000-0003-4172-4606}\inst{\ref{aff19}}
\and P.~B.~Lilje\orcid{0000-0003-4324-7794}\inst{\ref{aff53}}
\and V.~Lindholm\orcid{0000-0003-2317-5471}\inst{\ref{aff62},\ref{aff63}}
\and I.~Lloro\orcid{0000-0001-5966-1434}\inst{\ref{aff64}}
\and E.~Maiorano\orcid{0000-0003-2593-4355}\inst{\ref{aff17}}
\and O.~Mansutti\orcid{0000-0001-5758-4658}\inst{\ref{aff13}}
\and O.~Marggraf\orcid{0000-0001-7242-3852}\inst{\ref{aff65}}
\and K.~Markovic\orcid{0000-0001-6764-073X}\inst{\ref{aff54}}
\and M.~Martinelli\orcid{0000-0002-6943-7732}\inst{\ref{aff35},\ref{aff66}}
\and N.~Martinet\orcid{0000-0003-2786-7790}\inst{\ref{aff27}}
\and F.~Marulli\orcid{0000-0002-8850-0303}\inst{\ref{aff67},\ref{aff17},\ref{aff18}}
\and R.~Massey\orcid{0000-0002-6085-3780}\inst{\ref{aff68}}
\and E.~Medinaceli\orcid{0000-0002-4040-7783}\inst{\ref{aff17}}
\and M.~Melchior\inst{\ref{aff69}}
\and Y.~Mellier\inst{\ref{aff70},\ref{aff71}}
\and M.~Meneghetti\orcid{0000-0003-1225-7084}\inst{\ref{aff17},\ref{aff18}}
\and E.~Merlin\orcid{0000-0001-6870-8900}\inst{\ref{aff35}}
\and G.~Meylan\inst{\ref{aff72}}
\and M.~Moresco\orcid{0000-0002-7616-7136}\inst{\ref{aff67},\ref{aff17}}
\and L.~Moscardini\orcid{0000-0002-3473-6716}\inst{\ref{aff67},\ref{aff17},\ref{aff18}}
\and C.~Neissner\orcid{0000-0001-8524-4968}\inst{\ref{aff73},\ref{aff32}}
\and S.-M.~Niemi\inst{\ref{aff74}}
\and C.~Padilla\orcid{0000-0001-7951-0166}\inst{\ref{aff73}}
\and S.~Paltani\orcid{0000-0002-8108-9179}\inst{\ref{aff45}}
\and F.~Pasian\orcid{0000-0002-4869-3227}\inst{\ref{aff13}}
\and K.~Pedersen\inst{\ref{aff75}}
\and V.~Pettorino\inst{\ref{aff74}}
\and S.~Pires\orcid{0000-0002-0249-2104}\inst{\ref{aff49}}
\and G.~Polenta\orcid{0000-0003-4067-9196}\inst{\ref{aff76}}
\and M.~Poncet\inst{\ref{aff77}}
\and L.~A.~Popa\inst{\ref{aff78}}
\and F.~Raison\orcid{0000-0002-7819-6918}\inst{\ref{aff51}}
\and A.~Renzi\orcid{0000-0001-9856-1970}\inst{\ref{aff79},\ref{aff47}}
\and J.~Rhodes\orcid{0000-0002-4485-8549}\inst{\ref{aff54}}
\and G.~Riccio\inst{\ref{aff23}}
\and E.~Romelli\orcid{0000-0003-3069-9222}\inst{\ref{aff13}}
\and M.~Roncarelli\orcid{0000-0001-9587-7822}\inst{\ref{aff17}}
\and C.~Rosset\orcid{0000-0003-0286-2192}\inst{\ref{aff80}}
\and E.~Rossetti\orcid{0000-0003-0238-4047}\inst{\ref{aff16}}
\and R.~Saglia\orcid{0000-0003-0378-7032}\inst{\ref{aff52},\ref{aff51}}
\and Z.~Sakr\orcid{0000-0002-4823-3757}\inst{\ref{aff10},\ref{aff2},\ref{aff81}}
\and A.~G.~S\'anchez\orcid{0000-0003-1198-831X}\inst{\ref{aff51}}
\and D.~Sapone\orcid{0000-0001-7089-4503}\inst{\ref{aff82}}
\and B.~Sartoris\orcid{0000-0003-1337-5269}\inst{\ref{aff52},\ref{aff13}}
\and M.~Schirmer\orcid{0000-0003-2568-9994}\inst{\ref{aff59}}
\and P.~Schneider\orcid{0000-0001-8561-2679}\inst{\ref{aff65}}
\and T.~Schrabback\orcid{0000-0002-6987-7834}\inst{\ref{aff83}}
\and A.~Secroun\orcid{0000-0003-0505-3710}\inst{\ref{aff84}}
\and G.~Seidel\orcid{0000-0003-2907-353X}\inst{\ref{aff59}}
\and S.~Serrano\orcid{0000-0002-0211-2861}\inst{\ref{aff8},\ref{aff85},\ref{aff3}}
\and C.~Sirignano\orcid{0000-0002-0995-7146}\inst{\ref{aff79},\ref{aff47}}
\and G.~Sirri\orcid{0000-0003-2626-2853}\inst{\ref{aff18}}
\and L.~Stanco\orcid{0000-0002-9706-5104}\inst{\ref{aff47}}
\and J.~Steinwagner\orcid{0000-0001-7443-1047}\inst{\ref{aff51}}
\and P.~Tallada-Cresp\'{i}\orcid{0000-0002-1336-8328}\inst{\ref{aff31},\ref{aff32}}
\and I.~Tereno\inst{\ref{aff86},\ref{aff87}}
\and R.~Toledo-Moreo\orcid{0000-0002-2997-4859}\inst{\ref{aff88}}
\and F.~Torradeflot\orcid{0000-0003-1160-1517}\inst{\ref{aff32},\ref{aff31}}
\and L.~Valenziano\orcid{0000-0002-1170-0104}\inst{\ref{aff17},\ref{aff89}}
\and T.~Vassallo\orcid{0000-0001-6512-6358}\inst{\ref{aff52},\ref{aff13}}
\and Y.~Wang\orcid{0000-0002-4749-2984}\inst{\ref{aff90}}
\and J.~Weller\orcid{0000-0002-8282-2010}\inst{\ref{aff52},\ref{aff51}}
\and E.~Zucca\orcid{0000-0002-5845-8132}\inst{\ref{aff17}}
\and C.~Burigana\orcid{0000-0002-3005-5796}\inst{\ref{aff91},\ref{aff89}}
\and G.~Fabbian\orcid{0000-0002-3255-4695}\inst{\ref{aff92},\ref{aff93}}
\and F.~Finelli\orcid{0000-0002-6694-3269}\inst{\ref{aff17},\ref{aff89}}
\and A.~Pezzotta\orcid{0000-0003-0726-2268}\inst{\ref{aff51}}
\and V.~Scottez\inst{\ref{aff70},\ref{aff94}}
\and M.~Viel\orcid{0000-0002-2642-5707}\inst{\ref{aff12},\ref{aff13},\ref{aff15},\ref{aff14},\ref{aff95}}}
										   
\institute{Department of Astrophysics, University of Zurich, Winterthurerstrasse 190, 8057 Zurich, Switzerland\label{aff1}
\and
Institut de Recherche en Astrophysique et Plan\'etologie (IRAP), Universit\'e de Toulouse, CNRS, UPS, CNES, 14 Av. Edouard Belin, 31400 Toulouse, France\label{aff2}
\and
Institute of Space Sciences (ICE, CSIC), Campus UAB, Carrer de Can Magrans, s/n, 08193 Barcelona, Spain\label{aff3}
\and
Institut de Ciencies de l'Espai (IEEC-CSIC), Campus UAB, Carrer de Can Magrans, s/n Cerdanyola del Vall\'es, 08193 Barcelona, Spain\label{aff4}
\and
Laboratoire Univers et Th\'eorie, Observatoire de Paris, Universit\'e PSL, Universit\'e Paris Cit\'e, CNRS, 92190 Meudon, France\label{aff5}
\and
Kobayashi-Maskawa Institute for the Origin of Particles and the Universe, Nagoya University, Chikusa-ku, Nagoya, 464-8602, Japan\label{aff6}
\and
Institute for Advanced Research, Nagoya University, Chikusa-ku, Nagoya, 464-8601, Japan\label{aff7}
\and
Institut d'Estudis Espacials de Catalunya (IEEC),  Edifici RDIT, Campus UPC, 08860 Castelldefels, Barcelona, Spain\label{aff8}
\and
Universit\'e de Gen\`eve, D\'epartement de Physique Th\'eorique and Centre for Astroparticle Physics, 24 quai Ernest-Ansermet, CH-1211 Gen\`eve 4, Switzerland\label{aff9}
\and
Institut f\"ur Theoretische Physik, University of Heidelberg, Philosophenweg 16, 69120 Heidelberg, Germany\label{aff10}
\and
INAF-Osservatorio Astronomico di Brera, Via Brera 28, 20122 Milano, Italy\label{aff11}
\and
IFPU, Institute for Fundamental Physics of the Universe, via Beirut 2, 34151 Trieste, Italy\label{aff12}
\and
INAF-Osservatorio Astronomico di Trieste, Via G. B. Tiepolo 11, 34143 Trieste, Italy\label{aff13}
\and
INFN, Sezione di Trieste, Via Valerio 2, 34127 Trieste TS, Italy\label{aff14}
\and
SISSA, International School for Advanced Studies, Via Bonomea 265, 34136 Trieste TS, Italy\label{aff15}
\and
Dipartimento di Fisica e Astronomia, Universit\`a di Bologna, Via Gobetti 93/2, 40129 Bologna, Italy\label{aff16}
\and
INAF-Osservatorio di Astrofisica e Scienza dello Spazio di Bologna, Via Piero Gobetti 93/3, 40129 Bologna, Italy\label{aff17}
\and
INFN-Sezione di Bologna, Viale Berti Pichat 6/2, 40127 Bologna, Italy\label{aff18}
\and
INAF-Osservatorio Astrofisico di Torino, Via Osservatorio 20, 10025 Pino Torinese (TO), Italy\label{aff19}
\and
Dipartimento di Fisica, Universit\`a di Genova, Via Dodecaneso 33, 16146, Genova, Italy\label{aff20}
\and
INFN-Sezione di Genova, Via Dodecaneso 33, 16146, Genova, Italy\label{aff21}
\and
Department of Physics "E. Pancini", University Federico II, Via Cinthia 6, 80126, Napoli, Italy\label{aff22}
\and
INAF-Osservatorio Astronomico di Capodimonte, Via Moiariello 16, 80131 Napoli, Italy\label{aff23}
\and
INFN section of Naples, Via Cinthia 6, 80126, Napoli, Italy\label{aff24}
\and
Instituto de Astrof\'isica e Ci\^encias do Espa\c{c}o, Universidade do Porto, CAUP, Rua das Estrelas, PT4150-762 Porto, Portugal\label{aff25}
\and
Faculdade de Ci\^encias da Universidade do Porto, Rua do Campo de Alegre, 4150-007 Porto, Portugal\label{aff26}
\and
Aix-Marseille Universit\'e, CNRS, CNES, LAM, Marseille, France\label{aff27}
\and
Dipartimento di Fisica, Universit\`a degli Studi di Torino, Via P. Giuria 1, 10125 Torino, Italy\label{aff28}
\and
INFN-Sezione di Torino, Via P. Giuria 1, 10125 Torino, Italy\label{aff29}
\and
INAF-IASF Milano, Via Alfonso Corti 12, 20133 Milano, Italy\label{aff30}
\and
Centro de Investigaciones Energ\'eticas, Medioambientales y Tecnol\'ogicas (CIEMAT), Avenida Complutense 40, 28040 Madrid, Spain\label{aff31}
\and
Port d'Informaci\'{o} Cient\'{i}fica, Campus UAB, C. Albareda s/n, 08193 Bellaterra (Barcelona), Spain\label{aff32}
\and
Institute for Theoretical Particle Physics and Cosmology (TTK), RWTH Aachen University, 52056 Aachen, Germany\label{aff33}
\and
Institute of Cosmology and Gravitation, University of Portsmouth, Portsmouth PO1 3FX, UK\label{aff34}
\and
INAF-Osservatorio Astronomico di Roma, Via Frascati 33, 00078 Monteporzio Catone, Italy\label{aff35}
\and
Dipartimento di Fisica e Astronomia "Augusto Righi" - Alma Mater Studiorum Universit\`a di Bologna, Viale Berti Pichat 6/2, 40127 Bologna, Italy\label{aff36}
\and
Instituto de Astrof\'isica de Canarias, Calle V\'ia L\'actea s/n, 38204, San Crist\'obal de La Laguna, Tenerife, Spain\label{aff37}
\and
Institute for Astronomy, University of Edinburgh, Royal Observatory, Blackford Hill, Edinburgh EH9 3HJ, UK\label{aff38}
\and
European Space Agency/ESRIN, Largo Galileo Galilei 1, 00044 Frascati, Roma, Italy\label{aff39}
\and
ESAC/ESA, Camino Bajo del Castillo, s/n., Urb. Villafranca del Castillo, 28692 Villanueva de la Ca\~nada, Madrid, Spain\label{aff40}
\and
Universit\'e Claude Bernard Lyon 1, CNRS/IN2P3, IP2I Lyon, UMR 5822, Villeurbanne, F-69100, France\label{aff41}
\and
Institut de Ci\`{e}ncies del Cosmos (ICCUB), Universitat de Barcelona (IEEC-UB), Mart\'{i} i Franqu\`{e}s 1, 08028 Barcelona, Spain\label{aff42}
\and
Instituci\'o Catalana de Recerca i Estudis Avan\c{c}ats (ICREA), Passeig de Llu\'{\i}s Companys 23, 08010 Barcelona, Spain\label{aff43}
\and
UCB Lyon 1, CNRS/IN2P3, IUF, IP2I Lyon, 4 rue Enrico Fermi, 69622 Villeurbanne, France\label{aff44}
\and
Department of Astronomy, University of Geneva, ch. d'Ecogia 16, 1290 Versoix, Switzerland\label{aff45}
\and
Jodrell Bank Centre for Astrophysics, Department of Physics and Astronomy, University of Manchester, Oxford Road, Manchester M13 9PL, UK\label{aff46}
\and
INFN-Padova, Via Marzolo 8, 35131 Padova, Italy\label{aff47}
\and
INAF-Istituto di Astrofisica e Planetologia Spaziali, via del Fosso del Cavaliere, 100, 00100 Roma, Italy\label{aff48}
\and
Universit\'e Paris-Saclay, Universit\'e Paris Cit\'e, CEA, CNRS, AIM, 91191, Gif-sur-Yvette, France\label{aff49}
\and
INAF-Osservatorio Astronomico di Padova, Via dell'Osservatorio 5, 35122 Padova, Italy\label{aff50}
\and
Max Planck Institute for Extraterrestrial Physics, Giessenbachstr. 1, 85748 Garching, Germany\label{aff51}
\and
Universit\"ats-Sternwarte M\"unchen, Fakult\"at f\"ur Physik, Ludwig-Maximilians-Universit\"at M\"unchen, Scheinerstrasse 1, 81679 M\"unchen, Germany\label{aff52}
\and
Institute of Theoretical Astrophysics, University of Oslo, P.O. Box 1029 Blindern, 0315 Oslo, Norway\label{aff53}
\and
Jet Propulsion Laboratory, California Institute of Technology, 4800 Oak Grove Drive, Pasadena, CA, 91109, USA\label{aff54}
\and
Felix Hormuth Engineering, Goethestr. 17, 69181 Leimen, Germany\label{aff55}
\and
Technical University of Denmark, Elektrovej 327, 2800 Kgs. Lyngby, Denmark\label{aff56}
\and
Cosmic Dawn Center (DAWN), Denmark\label{aff57}
\and
Universit\'e Paris-Saclay, CNRS/IN2P3, IJCLab, 91405 Orsay, France\label{aff58}
\and
Max-Planck-Institut f\"ur Astronomie, K\"onigstuhl 17, 69117 Heidelberg, Germany\label{aff59}
\and
NASA Goddard Space Flight Center, Greenbelt, MD 20771, USA\label{aff60}
\and
Department of Physics and Helsinki Institute of Physics, Gustaf H\"allstr\"omin katu 2, 00014 University of Helsinki, Finland\label{aff61}
\and
Department of Physics, P.O. Box 64, 00014 University of Helsinki, Finland\label{aff62}
\and
Helsinki Institute of Physics, Gustaf H{\"a}llstr{\"o}min katu 2, University of Helsinki, Helsinki, Finland\label{aff63}
\and
NOVA optical infrared instrumentation group at ASTRON, Oude Hoogeveensedijk 4, 7991PD, Dwingeloo, The Netherlands\label{aff64}
\and
Universit\"at Bonn, Argelander-Institut f\"ur Astronomie, Auf dem H\"ugel 71, 53121 Bonn, Germany\label{aff65}
\and
INFN-Sezione di Roma, Piazzale Aldo Moro, 2 - c/o Dipartimento di Fisica, Edificio G. Marconi, 00185 Roma, Italy\label{aff66}
\and
Dipartimento di Fisica e Astronomia "Augusto Righi" - Alma Mater Studiorum Universit\`a di Bologna, via Piero Gobetti 93/2, 40129 Bologna, Italy\label{aff67}
\and
Department of Physics, Institute for Computational Cosmology, Durham University, South Road, DH1 3LE, UK\label{aff68}
\and
University of Applied Sciences and Arts of Northwestern Switzerland, School of Engineering, 5210 Windisch, Switzerland\label{aff69}
\and
Institut d'Astrophysique de Paris, 98bis Boulevard Arago, 75014, Paris, France\label{aff70}
\and
Institut d'Astrophysique de Paris, UMR 7095, CNRS, and Sorbonne Universit\'e, 98 bis boulevard Arago, 75014 Paris, France\label{aff71}
\and
Institute of Physics, Laboratory of Astrophysics, Ecole Polytechnique F\'ed\'erale de Lausanne (EPFL), Observatoire de Sauverny, 1290 Versoix, Switzerland\label{aff72}
\and
Institut de F\'{i}sica d'Altes Energies (IFAE), The Barcelona Institute of Science and Technology, Campus UAB, 08193 Bellaterra (Barcelona), Spain\label{aff73}
\and
European Space Agency/ESTEC, Keplerlaan 1, 2201 AZ Noordwijk, The Netherlands\label{aff74}
\and
DARK, Niels Bohr Institute, University of Copenhagen, Jagtvej 155, 2200 Copenhagen, Denmark\label{aff75}
\and
Space Science Data Center, Italian Space Agency, via del Politecnico snc, 00133 Roma, Italy\label{aff76}
\and
Centre National d'Etudes Spatiales -- Centre spatial de Toulouse, 18 avenue Edouard Belin, 31401 Toulouse Cedex 9, France\label{aff77}
\and
Institute of Space Science, Str. Atomistilor, nr. 409 M\u{a}gurele, Ilfov, 077125, Romania\label{aff78}
\and
Dipartimento di Fisica e Astronomia "G. Galilei", Universit\`a di Padova, Via Marzolo 8, 35131 Padova, Italy\label{aff79}
\and
Universit\'e Paris Cit\'e, CNRS, Astroparticule et Cosmologie, 75013 Paris, France\label{aff80}
\and
Universit\'e St Joseph; Faculty of Sciences, Beirut, Lebanon\label{aff81}
\and
Departamento de F\'isica, FCFM, Universidad de Chile, Blanco Encalada 2008, Santiago, Chile\label{aff82}
\and
Universit\"at Innsbruck, Institut f\"ur Astro- und Teilchenphysik, Technikerstr. 25/8, 6020 Innsbruck, Austria\label{aff83}
\and
Aix-Marseille Universit\'e, CNRS/IN2P3, CPPM, Marseille, France\label{aff84}
\and
Satlantis, University Science Park, Sede Bld 48940, Leioa-Bilbao, Spain\label{aff85}
\and
Departamento de F\'isica, Faculdade de Ci\^encias, Universidade de Lisboa, Edif\'icio C8, Campo Grande, PT1749-016 Lisboa, Portugal\label{aff86}
\and
Instituto de Astrof\'isica e Ci\^encias do Espa\c{c}o, Faculdade de Ci\^encias, Universidade de Lisboa, Tapada da Ajuda, 1349-018 Lisboa, Portugal\label{aff87}
\and
Universidad Polit\'ecnica de Cartagena, Departamento de Electr\'onica y Tecnolog\'ia de Computadoras,  Plaza del Hospital 1, 30202 Cartagena, Spain\label{aff88}
\and
INFN-Bologna, Via Irnerio 46, 40126 Bologna, Italy\label{aff89}
\and
Infrared Processing and Analysis Center, California Institute of Technology, Pasadena, CA 91125, USA\label{aff90}
\and
INAF, Istituto di Radioastronomia, Via Piero Gobetti 101, 40129 Bologna, Italy\label{aff91}
\and
Kavli Institute for Cosmology Cambridge, Madingley Road, Cambridge, CB3 0HA, UK\label{aff92}
\and
Institute of Astronomy, University of Cambridge, Madingley Road, Cambridge CB3 0HA, UK\label{aff93}
\and
Junia, EPA department, 41 Bd Vauban, 59800 Lille, France\label{aff94}
\and
ICSC - Centro Nazionale di Ricerca in High Performance Computing, Big Data e Quantum Computing, Via Magnanelli 2, Bologna, Italy\label{aff95}}    

%
%
%
%


%
%
\abstract{
Gravitational redshift and Doppler effects give rise to an antisymmetric component of the galaxy correlation function when cross-correlating two galaxy populations or two different tracers. In this paper, we assess the detectability of these effects in the \Euclid spectroscopic galaxy survey. We model the impact of gravitational redshift on the observed redshift of galaxies in the Flagship mock catalogue using a Navarro--Frenk--White profile for the host haloes. We isolate these relativistic effects, largely subdominant in the standard analysis, by splitting the galaxy catalogue into two populations of faint and bright objects and estimating the dipole of their cross-correlation in four redshift bins.

In the simulated catalogue, we detect the dipole signal on scales below $ 30\,h^{-1}{\rm Mpc}$, with detection significances of $4\,\sigma$ and $3\,\sigma$ in the two lowest redshift bins, respectively. At higher redshifts, the detection significance drops below $2\,\sigma$. Overall, we estimate the total detection significance in the \Euclid spectroscopic sample to be approximately $6\,\sigma$. We find that on small scales, the major contribution to the signal comes from the nonlinear gravitational potential. Our study on the Flagship mock catalogue shows that this observable can be detected in Euclid Data Release 2 and beyond.
    }
%
%
%
%
   \titlerunning{\Euclid: Relativistic effects in the dipole of the two-point correlation function}
   \authorrunning{F. Lepori et al.}
   
   \maketitle
%
%
%
%
   
\section{\label{sc:Intro}Introduction}

The current picture of the Universe and its evolution history is based on the concordance $\Lambda$ cold dark matter ($\Lambda$CDM) model, which describes gravitational phenomena according to the laws of general relativity and assumes the Universe to be statistically homogeneous and isotropic on cosmological scales.
This simple model is capable of fitting a wide variety of observations:
the temperature and polarisation anisotropies of the cosmic microwave background~\citep{wmap, Plank2020}, the clustering of galaxies on large scales~\citep{eBOSS, DESI:2024mwx}, weak gravitational lensing, which causes distortion in the observed shapes of galaxies~\citep{DESY3}, and the distance measurements from supernovae~\citep{SN-Scolnic, Abbott:2024}.
However, the standard model is not fully satisfactory. From a theoretical point of view, it requires that the majority of the energy content in our Universe takes some unknown form: a cosmological constant, whose
inferred value lacks fundamental understanding, drives the current accelerated expansion of the Universe, while the matter content is mostly composed of a non-baryonic component that is yet to be directly detected. 
Furthermore, as recent cosmological experiments have become more precise, some tensions between parameters inferred from different probes have emerged~\citep{DiValentino:2020vvd,DiValentino:2020zio, Schoneberg:2021qvd, DESI:2024mwx,Tutusaus:2023aux}. Currently, it remains unclear whether these discrepancies either reveal a breakdown of the $\Lambda$CDM model, a sign that astrophysical or systematic effects have been overlooked in the analyses in tension, or whether they are nothing more than a statistical fluke.

One of the main goals of the next generation of cosmological surveys is to resolve these issues, in particular by testing one of the pillars of the standard model: Einstein's theory of general relativity. 
In the near future, the Euclid Survey~\citep{Laureijs:2011gra, Amendola:2016saw, EuclidSkyOverview} is set to play a key role in this challenge. By using two complementary probes -- weak lensing (including galaxy clustering and galaxy-galaxy lensing from its photometric survey) and 3D galaxy clustering (from its spectroscopic survey) -- \Euclid will map the expansion history of the Universe as well as the growth of cosmic structure within it.

The Euclid Spectroscopic Survey will provide a catalogue of more than 25 million emission-line galaxies using slitless spectroscopy for redshift determination, in the redshift range between $z= 0.9$ and $z=1.8$. 
One of the key measurements that \Euclid will carry out from these data is that of the growth rate at different cosmic times, using redshift-space distortions~\citep[RSDs;][]{Kaiser1987}. RSDs manifest because we do not directly observe galaxies where they are but rather infer their physical distances from their observed redshifts, assuming that the latter is approximately given by the Hubble expansion. However, the observed redshift is also affected by the peculiar velocities of galaxies through the Doppler shift, for example.
Since we do not know the peculiar velocities of individual objects, we cannot correct each redshift individually but rather model the distortions induced by the peculiar motions in the two-point statistics. 
The main effect is that galaxies moving along the line-of-sight toward an overdensity will appear closer to each other.
Therefore, the observed two-point correlation function (or its Fourier counterpart, the power spectrum) is not isotropic, but also includes even multipoles
(quadrupole and hexadecapole) that can be measured to extract cosmological information~\citep{Alam:2015rsa, BOSS:2016off}. 

More recently, a fully relativistic computation has shown that in addition to the Kaiser RSDs, typically exploited in anisotropic galaxy clustering analysis, there are other subtle physical effects that can also influence the observed distribution of galaxies, such as gravitational redshift, Doppler effects, Shapiro time delay, and the integrated Sachs--Wolfe effect~\citep{Yoo:2009au, Yoo:2010ni, Bonvin:2011bg, Challinor:2011bk,Jeong:2011as}.
Although their impact on even multipoles of the correlation function has
been shown to be negligible even for future galaxy surveys~\citep{Jelic-Cizmek:2020pkh}, it is possible to isolate them using multiple tracers. Importantly, while standard RSDs generate only even multipoles of the correlation function at leading order, the gravitational redshift and further Doppler-type contributions generate odd multipoles in the two-point cross-correlation function~\citep{McDonald:2009ud, Bonvin:2013ogt}.

Measuring the odd multipoles is interesting, as they provide complementary information to the even multipoles. In particular, the dipole is sourced in part by the gravitational redshift, so in principle is sensitive to the distortion of time~\citep{Sobral-Blanco:2022oel}. Measuring the distortion of time through the dipole, in combination with existing probes, would thus allow for a range of novel tests. These include: a test of the validity of the equivalence principle for dark matter, by combining with RSDs~\citep{Bonvin:2018ckp}, 
a test of the validity of local position invariance~\citep{Saga:2021osv}, a novel way to measure the anisotropic stress, by combining with weak gravitational lensing~\citep{Sobral-Blanco:2021cks,Tutusaus:2022cab}, 
and more generally a way to distinguish between modifications of gravity and additional forces in the dark sector~\citep{Castello:2022uuu, Bonvin:2022tii, Castello:2023zjr}. Furthermore, the dipole and octupole can be used to improve our understanding of astrophysical properties of galaxies~\citep{Sobral-Blanco:2024qlb}.

Gravitational redshift has been measured in galaxy clusters~\citep{Wojtak_2011, Sadeh:2014rya, Rosselli:2022qoz}, 
and also in large-scale structure, with a $2.8\,\sigma$ detection of the relativistic dipole  reported at scales below $10\,h^{-1}\,\mathrm{Mpc}$ using data from SDSS-III~\citep{Alam:2017izi}. The detectability of the dipole has been investigated for different combinations of tracers and future cosmological surveys using analytical models for both signals and covariance, showing that it will be robustly detected in the linear and nonlinear
regime, see for example~\cite{Bonvin:2015kuc}, ~\cite{Hall:2016bmm}, ~\cite{Lepori:2017twd}, ~\cite{Bonvin:2018ckp}, ~\cite{Lepori:2019cqp} and ~\cite{Saga:2021jrh}.
The dipole has moreover been detected in $N$-body simulations, cross-correlating halo populations with different masses~\citep{Breton:2018wzk, Beutler:2020evf}, and galaxy populations at low redshift selected based on their apparent flux~\citep{Bonvin:2023jjq}. 

In this paper, we aim to provide a realistic forecast on the detectability of the relativistic dipole in the Euclid Spectroscopic Galaxy Catalogue. Our forecasts are based on a mock catalogue constructed from the Flagship simulation. 
In~\cref{sec:2-theory} we present the theoretical basis of our work and discuss the models for the dipole in the linear and nonlinear regime. In~\cref{sec:3-method} we describe the selection we have applied to the Flagship mock catalogue and the methodology of our analysis. In~\cref{sec:4-results} we present the main results of our paper: we compare the measurements of the dipole for different selections, identify the main contributions to the measured dipole, and assess the detection significance in our mock. 
Finally, we draw the conclusions of our work in~\cref{sec:conc} and further discuss the implications  for the measurements in the planned data releases of \Euclid. 

\section{Relativistic effects in galaxy clustering}
\label{sec:2-theory}

A galaxy survey can be used to measure the statistics of fluctuations in the galaxy counts,
\be
\De(\vec{n},z)  =\frac{N(\vec{n},z)-\bar N(z)}{\bar N(z)} \,, \label{eq:nc}
\ee
where $N(\vec{n},z)$ is the number of objects observed in a pixel of fixed solid angle in direction $\vec{n}$ at redshift $z$, and
${\bar N(z)}$ is the average count over all pixels.  
The observed angular positions and redshift of galaxies are influenced by the fact that we do not live in a perfectly homogeneous and isotropic Universe. Consequently, the wavelengths and paths of photons are affected by perturbations, which in turn affects the observed galaxy counts through RSDs and other relativistic effects. 
We represent the spacetime metric using the metric of a spatially flat Friedmann--Lemaître--Robertson--Walker (FLRW) universe, with scalar perturbations given by the peculiar gravitational potentials $\Phi$ and $\Psi$.
The perturbed line element in longitudinal gauge is 
\be
\de s^2 = a^2(\eta)\left[-\left(1+\frac{2\Psi}{c^2}\right)\,c^2\de\eta^2+ \left(1-\frac{2\Phi}{c^2}\right) 
\,\delta_{ij} \,\de x^i\de x^j\right] \,, 
\label{eq:met}
\ee
where $c$ is the speed of light, $\eta$ denotes the conformal time, $a$ represents the cosmological scale factor,
$x^i$ are comoving Cartesian coordinates, and $\delta_{ij}$ is the Kronecker delta operator.\footnote{We adopt Einstein's summation convention in which repeated indices are summed over.}

The fluctuations in galaxy counts have been computed in a perturbed spacetime assuming linear perturbations in~\citet{Yoo:2009au},~\citet{Yoo:2010ni}, ~\citet{Bonvin:2011bg},~\citet{Challinor:2011bk}, and ~\citet{Jeong:2011as},
and to higher orders in perturbation theory in~\citet{Yoo:2014sfa},~\citet{Bertacca:2014dra}, ~\citet{DiDio:2014lka},~\citet{Nielsen:2016ldx},~\citet{DiDio:2018zmk}, and~\citet{DiDio:2020jvo}. 
The leading order contributions relevant for this work are: 
\begin{multline}
\Delta(\vec{n},z)=b\,\delta-\frac{1}{\mathcal{H}}\partial_r(\vec{V}\cdot\vec{n})  \label{eq:Delta} \\
+\frac{1}{c\,\mathcal{H}}\partial_r\Psi 
+\left(1-\frac{\mathcal{H}^\prime}{\mathcal{H}^2}+\frac{(5s-2)\,c}{r\mathcal{H}} -5s+f^{\rm evo}\right)\frac{\vec{V}}{c}\cdot\vec{n}+\frac{1}{\mathcal{H}}\frac{\vec{V}^\prime}{c}\cdot\vec{n}\,.
\end{multline}
In this expression, $\delta$ is the density contrast in the comoving gauge, $\mathbf{V}$ is the peculiar velocity at the source in longitudinal gauge, $r \equiv r(z)$ is the comoving distance corresponding to $z$, $\mathcal{H}\equiv a^{\prime}/a $ is the conformal Hubble parameter, and the `prime' symbol represents a derivative with respect to the conformal time. 
The redshift-dependent coefficients $b$, $s$, and $f^\mathrm{evo}$ are the linear galaxy bias, the local count slope, and the evolution bias, respectively.   
The first line in Eq.~\eqref{eq:Delta} contains the dominant contributions to the fluctuation of the galaxy counts due to the linear density fluctuation of objects and the RSDs term. 
The second line represents the subdominant relativistic contributions and includes
gravitational redshift (the first term in the second line) and Doppler terms, which we aim to measure with \Euclid.

Equation~\eqref{eq:Delta} does not include lensing magnification, nor 
subdominant relativistic effects that are proportional to the gravitational potentials -- 
the full expression can be found in ~\cite{Challinor:2011bk}, \cite{Bonvin:2011bg}, and~\cite{DiDio:2013bqa}. 
These extra relativistic effects are suppressed compared to the gravitational redshift and Doppler corrections, and therefore can be safely neglected~\citep{Bonvin:2014owa}. Lensing magnification also contributes to the number counts, and is expected to be an important effect for both the photometric and the spectroscopic Euclid Surveys~\citep{Lepori-EP19, Euclid:2023qyw}, and the cross-correlation of photometric galaxy clustering with the cosmic microwave background~\citep{Bermejo-Climent:2019spz}. 
However, the lensing contribution to the dipole is negligible for most applications, because the only contribution comes from an integral over the correlation scale.~\cite{Bonvin:2013ogt} show that in linear theory the impact of lensing on the dipole is largely subdominant with respect to the Doppler and gravitational redshift at $z = 2$; this is confirmed to be valid in the nonlinear regime, using $N$-body simulations, in \cite{Breton:2018wzk}. 

Equation~\eqref{eq:Delta} assumes that the photon moves on null geodesics in a perturbed FLRW spacetime, without making any assumption on the dynamics of the perturbations. However, it can be further simplified assuming that matter obeys Euler's equation.\footnote{This assumption is violated in presence of  a fifth force acting on the dark matter component; see, for example,~\cite{Bonvin:2018ckp}.}
Using 
\begin{equation}
\vec{V}^\prime\cdot\vec{n} + \mathcal{H} \vec{V}\cdot\vec{n} + \partial_r\Psi  = 0\,, 
\end{equation}
Eq.~\eqref{eq:Delta} becomes
\begin{align}
\Delta(\vec{n},z)&= b\,\delta-\frac{1}{\mathcal{H}}\partial_r(\vec{V}\cdot\vec{n})  \label{eq:Delta2}  \\
&\quad  -\left(\frac{\mathcal{H}^\prime}{\mathcal{H}^2}-\frac{(5s-2)\,c}{r\mathcal{H}} + 5s - f^{\rm evo}\right)\frac{\vec{V}}{c}\cdot\vec{n}\,. \nonumber
\end{align}

\subsection{Relativistic dipole: Linear regime}

The information contained in the galaxy counts is normally compressed into summary statistics. For a spectroscopic survey, the most common data compression consists of estimating the multipoles of the two-point correlation function (2PCF) or the multipoles of the power spectrum in Fourier space. 
In real space, where only the galaxy overdensity contributes to the galaxy counts, the correlation function is isotropic as a consequence of the cosmological principle. 
The second term in the first line of Eq.~\eqref{eq:Delta}, known as Kaiser RSDs~\citep{Kaiser1987},
breaks the isotropy of the correlation function. At leading order, RSDs generate only
even multipoles of the correlation function: quadrupole and hexadecapole.
The relativistic effects in the second line of Eq.~\eqref{eq:Delta} or Eq.~\eqref{eq:Delta2} break the symmetry of the correlation function when two tracers with different biases are cross-correlated~\citep{McDonald:2009ud, Bonvin:2013ogt}. 
Therefore, by measuring the odd multipoles in the cross-correlation function, we can isolate the relativistic effects from the dominant density and Kaiser RSDs contributions. The main contribution
to the antisymmetric part of the cross-correlation function is given by the dipole and is the focus of this work.

To extract this relativistic dipole from the Euclid Spectroscopic Galaxy Catalogue, we need to split the sample into two distinct populations. Here we split it into a bright population and a faint population. 
This choice is straightforward to apply from an observational perspective, as flux is a directly observable quantity. Furthermore, assuming that galaxy luminosity traces mass, we expect the two populations to exhibit a significant bias difference for this selection, leading to a stronger signal, see for example~\cite{Bonvin:2013ogt} and~\cite{Alam:2017izi}.
The method for performing this split is described in~\cref{sec:3-method}, while in this section we discuss the theoretical modelling of this observable.
We denote the galaxy count fluctuations for the bright and faint populations by $\Delta_{\rm B}$ and $\Delta_{\rm F}$, respectively. 
In the wide-angle regime, the cross-correlation of the number counts between the two populations depends on three coordinates and is given by
\begin{equation}
\xi(z_1, z_2, \theta)= \langle \Delta_{\rm B}(\vec{n}_1, z_1)\Delta_{\rm F}(\vec{n}_2, z_2) \rangle \, ,
\end{equation}
where $\langle...\rangle$ denotes the ensemble average, and $\cos{\theta} = \vec{n}_1 \cdot \vec{n}_2$. 
A commonly used coordinate system is given by the mean redshift $\bar{z} = (z_1+z_2)/2$,  the separation between the two tracers $d$, estimated assuming a fiducial cosmology, and the orientation of the two tracers $\mu =  \vec{n} \cdot \hat{\vec{d}}$, where 
$\vec{n}$ and $\hat{\vec{d}}$ are unit vectors, with $\vec{n}$ pointing in
the direction of the midpoint of the two tracers and $\hat{\vec{d}}$ pointing from the bright to the faint tracer.
In this coordinate system, due to azimuthal symmetry around the direction $\vec{n}$, we can expand the correlation function into Legendre polynomials $L_\ell$:
\begin{equation}
   \xi(\bar{z}, d, \mu) = \sum_\ell \xi_\ell(\bar{z}, d)\,L_\ell(\mu)\,.
\end{equation}
The coefficients of this expansion $\xi_\ell(\bar{z}, d)$ are the multipoles of the correlation function. For separations $d$ much smaller than the comoving distance to the mean redshift $r(\bar{z})$, 
we can expand the dipole in powers of $d/r$, keeping only the leading-order terms~\citep{Bonvin:2013ogt}. 
The dominant contributions to the dipole are given by the relativistic dipole in the flat-sky approximation and the leading-order wide-angle contribution from the $d/r$ expansion,
\begin{equation}
  \xi_1(\bar{z}, d)  = \mathcal{R}^{\rm rel}(\bar{z})\,\nu_1(\bar{z}, d)  +  \mathcal{R}^{\rm wa}(\bar{z})\,
  \mu_2(\bar{z}, d)\,, \label{eq:dip-lin}
\end{equation}
where $\nu_1$ and $\mu_2$ are the following dimensionless integrals of the linear matter power spectrum
$P(\bar{z}, k)$,
\begin{align}
&\nu_1(\bar{z}, d)=\frac{1}{2\pi^2} \frac{\HH_0}{c}\int \diff  k\,k\,P(\bar{z}, k)\,j_1(kd)\, , \label{eq:nu1} \\
&\mu_2(\bar{z}, d)=\frac{1}{2\pi^2}\int \diff  k\, k^2\,P(\bar{z}, k)\,j_2(kd)\, ,\label{eq:mu2}
\end{align}
and $\mathcal{R}^{\rm rel}$ and  $\mathcal{R}^{\rm wa}$ are given by the redshift-dependent coefficients
\begin{align}
&\mathcal{R}^{\rm rel} = \frac{\HH}{\HH_0}  \,\left[\left(b_{\rm B}\, \mathcal{C}_{\rm F} - b_{\rm F}\,\mathcal{C}_{\rm B}\right)f + \frac{3}{5} \left(\mathcal{C}_{\rm F} - \mathcal{C}_{\rm B}\right)\,f^2\right]\,, \label{eq:Rrel} \\
& \mathcal{R}^{\rm wa} = -\frac{2}{5} \left( b_{\rm B}-b_{\rm F}\right) f\,\frac{d}{r}\, , \label{eq:Rwa} \\
& \mathcal{C}_{\rm B/F} = \frac{\HH^{\prime}}{\HH^2} + \frac{(2-5s_{\rm B/F})\,c}{r\,\HH} +5 s_{\rm B/F}  - f^{\rm evo}_{\rm B/F} \,,
\end{align}
with $f$ denoting the growth rate.

\subsection{Relativistic dipole: Nonlinear regime}
\label{sec:model-nl}

On small scales, it has been shown that the main contribution to the dipole comes from the gravitational redshift arising from the deep potential well of haloes at the galaxy position~\citep{Breton:2018wzk}. 
To describe the small-scale dipole signal from the halo potential, we adopt the phenomenological model presented in~\citet{Saga:2020tqb, Saga:2021jrh}, where, on top of the standard RSDs term and linear gravitational redshift term, the observed redshift is further modulated by $- (1+z)\, \overline{\Psi}^{0}/c^{2}$ with $\overline{\Psi}^{0}$ being the nonlinear gravitational potential at the galaxy's position, which we cannot simply characterise in a perturbative way. Although this correction gives a negligible contribution to even multipoles relative to the standard Doppler effect, it gives a major contribution to the dipole on small scales. 

Assuming that we cross-correlate two populations having a typical potential depth $\overline{\Psi}^{0}_{\rm B}$ for the bright population and $\overline{\Psi}^{0}_{\rm F}$ for the faint population, the full dipole signal is modelled as a sum of a linear part, given by Eq.~\eqref{eq:dip-lin}, and a nonlinear contribution,
\begin{equation}
 \xi^{\rm NL}_1(\bar{z}, d)  =  \xi_1(\bar{z}, d) + 
 \xi^{\Psi}_1(\bar{z}, d)\, . 
 \label{eq:dip-nlphi}
\end{equation}
Here, the second term on the right-hand side, $\xi^{\Psi}_1(\bar{z}, d)$, represents a nonlinear contribution to the dipole arising from the nonlinear potential, explicitly given by
\begin{equation}
\xi^{\Psi}_1(\bar{z}, d) = \mathcal{R}^{\Psi}(\bar{z})\, \nu_{\Psi} (\bar{z}, d)\, ,
\label{eq:xi-NL}
\end{equation}
where the functions $\nu_{\Psi}$ and $\mathcal{R}^{\Psi}$ are defined as
\begin{align}
& \nu_{\Psi} (\bar{z}, d) = \frac{1}{2\pi^2} \frac{c}{\HH} \int \diff  k\,k^3\,P(\bar{z}, k)\,j_1(kd)\, , \label{eq:nu-NL} \\
& \mathcal{R}^{\Psi}(\bar{z}) = -\left(\frac{\overline{\Psi}^0_{\rm B}}{c^2} - \frac{\overline{\Psi}^0_{\rm F}}{c^2} \right)\,\left[b_{\rm B} b_{\rm F} + \frac{3}{5} f \left( b_{\rm B} + b_{\rm F} \right) + \frac{3}{7} f^2  \right]. 
\end{align}
This expression shows that the dipole signal from the nonlinear potential is nonvanishing only when we cross-correlate different populations with $\overline{\Psi}^{0}_{\rm B} \neq \overline{\Psi}^{0}_{\rm F}$.

Together with the model of the nonlinear potential presented in \cref{sec:imple-grav}, the model in Eq.~\eqref{eq:dip-nlphi} has been shown to reproduce very well the measured dipole in $N$-body simulations down to $d\approx 5\,h^{-1}\,{\rm Mpc}$~\citep{Saga:2020tqb}.
Since the contribution from the nonlinear gravitational redshift given in Eq.~\eqref{eq:xi-NL} is suppressed as the separation $d$ increases, the total signal in Eq.~\eqref{eq:dip-nlphi} recovers the results of linear theory on large scales.

\begin{figure}[t!] 
    \centering
    \includegraphics[width=\columnwidth]{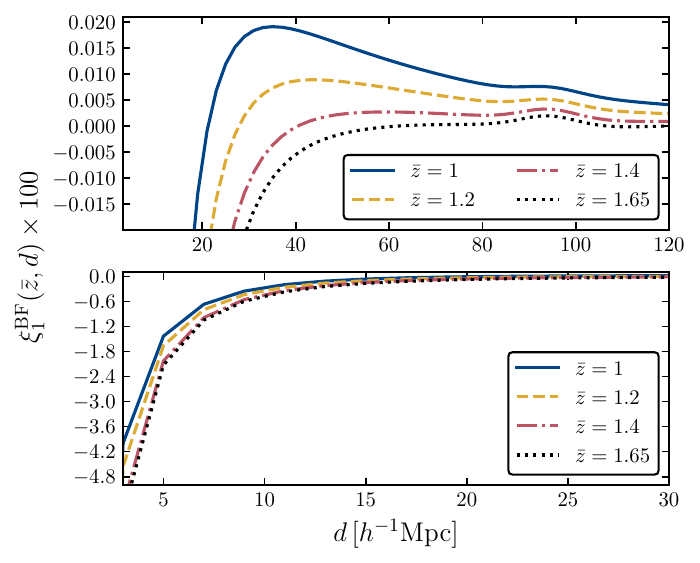}
    \caption{Dipole model, including the contribution from nonlinear gravitational redshift. The top panel includes linear scales up to $120\,h^{-1}\,{\rm Mpc}$, while the bottom panel is a zoom-in of the observable on nonlinear scales, up to $30\,h^{-1}\,{\rm Mpc}$. 
    In the two panels, we use a different range to display the dipole model for better visualization: the dipole amplitude is significantly larger on small scales.
    Specifics are taken from~\cref{tab:spec}, using the split $90\%$ faint, $10\%$ bright. Top and bottom panels focus on the large-scale and small-scale dipole predictions, respectively.}
    \label{fig:theory-dip}
\end{figure}

In~\cref{fig:theory-dip} we show the nonlinear model for the dipole at the redshifts considered in this work. The signal is largely boosted on scales below $30\,h^{-1}\,{\rm Mpc}$ by the contribution of the nonlinear gravitational redshift. 
In our model, density and velocity fields are assumed to be linear. We have tested the impact of nonlinearities in the matter field by replacing the linear matter power spectrum in Eq.~\eqref{eq:nu-NL} with the nonlinear power spectrum. In all scales of interest for our work, the difference between these two cases is negligible. 

Other models have been proposed in the literature, which use perturbation theory and include higher-order bias correction~\citep{DiDio:2018zmk, DiDio:2020jvo, Beutler:2020evf}. The main disadvantage of this perturbative approach is that it introduces extra parameters into the model that cannot be directly measured in simulations but must instead be fitted to the simulated data vectors. The model presented in~\cite{Saga:2021jrh} and outlined in this section is more predictive, primarily because the extra ingredients $\overline{\Psi}^{0}_{\rm B}$ and $\overline{\Psi}^{0}_{\rm F}$ can be estimated from simulations.

\subsection{Theory covariance}
\label{sec:cov}

The theory covariance for the linear dipole was first derived in \cite{Hall:2016bmm}. 
For the cross-correlation between two galaxy populations, there are three contributions: a pure shot noise term, a
pure sample variance term, and a shot noise--sample variance cross-term:
\begin{equation}
{\rm Cov}^{(\rm th)} = {\rm Cov}^{(\rm noise)} +
{\rm Cov}^{(\rm SV)} + {\rm Cov}^{({\rm SV}\,\times\,{\rm noise})}\,. \label{eq:cov-theory}
\end{equation}
The shot noise term is given by
\begin{equation}
{\rm Cov}^{(\rm noise)}_{ij} = \frac{3}{4\pi V \bar{n}_{\rm B} \bar{n}_{\rm F}\,\Delta d\,d_i^2}\delta_{ij}\,,
\end{equation}
the sample variance term can be written as
\begin{multline}
{\rm Cov}^{(\rm SV)}_{ij} = \frac{9}{4V}\left(\frac{\HH}{\HH_0}\right)^2\Bigg[\frac{2}{5} (b_{\rm F} \mathcal{C}_{\rm B} - b_{\rm B} \mathcal{C}_{\rm F})^2  \\
+ \frac{4}{7} (\mathcal{C}_{\rm B} - \mathcal{C}_{\rm F}) (b_{\rm F} \mathcal{C}_{\rm B} - b_{\rm B} \mathcal{C}_{\rm F}) f + \frac{2}{9} (\mathcal{C}_{\rm B} - \mathcal{C}_{\rm F})^2 f^2\Bigg]J_{ij}\, ,
\end{multline}
and the cross-term is
\begin{multline}
 {\rm Cov}^{({\rm SV}\,\times\,{\rm noise})}_{ij} = \frac{9}{2\, V}\!\left[\frac{1}{\bar{n}_{\rm F}}\!\left(\frac{b_{\rm B}^2}{3}+\frac{2b_{\rm B} f}{5}+\frac{f^2}{7} \right) \right. \\
 \left.+\frac{1}{\bar{n}_{\rm B}}\!\left(\frac{b_{\rm F}^2}{3}+\frac{2b_{\rm F} f}{5}+\frac{f^2}{7} \right) \right]G_{ij}\,,
\end{multline}
where $V$ denotes the volume of the redshift bin, $\bar{n}_{\rm B/F}$ is the average number density inside the bin, and $\Delta d$ is the bin width used to estimate the correlation function.
Here $G_{ij}$ and $J_{ij}$ are given by the following integrals of the linear matter power spectrum
\begin{align}
G_{ij}=&\frac{1}{\pi^2}\int \diff k \; k^2 P(\bar{z}, k)j_1(kd_i)j_1(kd_j)\, ,\\
J_{ij}=&\frac{1}{\pi^2} \frac{\HH_0^2}{c^2} \int \diff k \; P^2(\bar{z}, k)j_1(kd_i)j_1(kd_j)\,. 
\end{align}

\begin{figure}[t!] 
    \centering
    \includegraphics[width=\columnwidth]{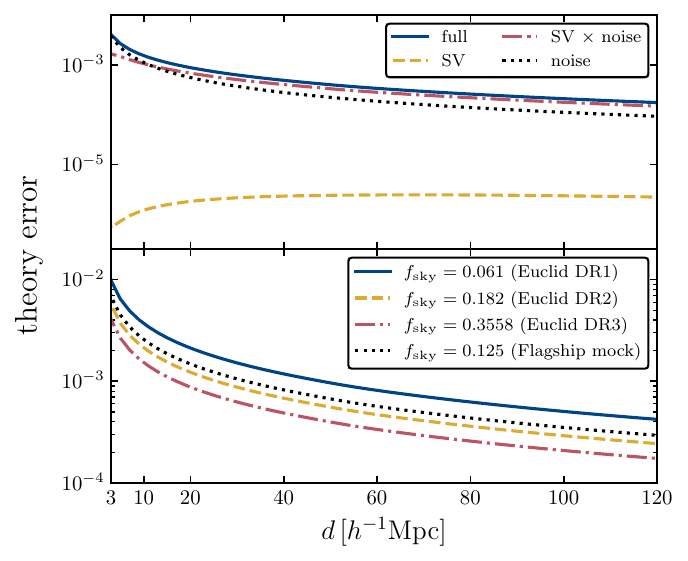}
    \caption{
    Theory uncertainty, given by the square root of the diagonal elements of the theory covariance, estimated for the redshift bin $\bar{z} = 1$, with specifics from~\cref{tab:spec}, split `$90\%$ faint, $10\%$ bright'. We have assumed a bin size of $\Delta d = 2\,h^{-1}\,{\rm Mpc}$. The top panel shows the different contributions to the uncertainty, 
    coming from the pure shot noise term (black dotted line), the purely sample variance term (yellow dashed line), and the cross-term (red dot-dashed line). 
    The sum of all contributions is shown as solid blue line. The bottom panel shows the theory uncertainty, 
    including all terms, for different sky coverages
    corresponding to the planned data releases of \Euclid and the Flagship mock used in this work. 
    }
    \label{fig:theory-cov}
\end{figure}

In the upper panel of~\cref{fig:theory-cov} we show the different contributions to the uncertainty
estimated from the linear covariance by taking the square root of its diagonal elements at $\bar{z} = 1$. 
The sample variance term ${(\rm SV)}$ is largely subdominant on all scales. This is due to the fact that density and RSDs do not contribute to it -- only relativistic effects affect this term. 
On scales larger than $15\,h^{-1}\,{\rm Mpc}$, the uncertainty 
is mostly due to the cross term $({\rm SV}\,\times\,{\rm noise})$, which, contrary to the sample variance term, is affected by density and RSDs. In the highly nonlinear regime, $d < 10\,h^{-1}\,{\rm Mpc}$, the Poisson noise contribution dominates over the others. 
As shown in~\cite{Saga:2021jrh}, the theory covariance for the nonlinear model presented in~\cref{sec:model-nl}
contains additional contributions from nonlinear gravitational redshift. However, they only affect the sample variance contributions
and turn out to be negligible. 

In the lower panel of~\cref{fig:theory-cov} we show how the theory uncertainty
scales with respect to the sky coverage, assuming 
a fixed number density of galaxies. The theory uncertainty
estimated for the Flagship mock is close to the uncertainty 
expected for the second data release of Euclid (DR2). We expect the uncertainty 
in Euclid DR3 to be reduced by about a factor $3/5$ compared to the Flagship mock.

\section{Method}
\label{sec:3-method}

\subsection{Simulated data}

\subsubsection{The Flagship galaxy mock}

The \Euclid Flagship simulation (FS hereafter) is the reference simulation for the Euclid mission and it has been used as a key ingredient for the scientific preparation of the mission. The FS features a simulation box of 3600\,$h^{-1}$\,Mpc on a side with $16\,000^3$ particles, leading to a mass resolution of $m_{\rm p} = 10^9\,h^{-1}\,M_{\odot}$. This 4 trillion particle simulation is one of the largest $N$-body simulation performed to date and meets the basic science requirements of the mission: it allows us to include the faintest galaxies that \Euclid will observe while sampling a cosmological volume comparable to what the Euclid Surveys will cover. The simulation is performed using \texttt{PKDGRAV3} \citep{Potter:16} on the `Piz Daint' supercomputer at the Swiss National Supercomputing Centre (CSCS). The FS uses the following values for the density parameters: $\Omega_{\rm m} = 0.319$, $\Omega_{\rm b} = 0.049$, and $\Omega_{\Lambda} = 0.681 - \Omega_{\rm rad} - \Omega_{\nu}$, with a radiation density $\Omega_{\rm rad} = 0.00005509$, and a contribution from massive neutrinos $\Omega_{\nu} = 0.00140343$. Additional parameters are: the dark energy equation-of-state parameter $w = -1.0$, the reduced Hubble constant $h = 0.67$, the scalar spectral index of initial fluctuations $n_{\rm s} = 0.96$, and the amplitude of the scalar power spectrum $A_{\rm s} = 2.1 \times 10^{-9}$ (corresponding to $\sigma_8 = 0.813$) at $k = 0.05\,{\rm Mpc}^{-1}$. The initial conditions are set at $z = 99$ using first-order Lagrangian perturbation theory (1LPT) displacements from a uniform particle grid. The transfer functions for the density field and the velocity field are generated at this initial redshift by {\sc class} \citep{CLASS} and \texttt{CONCEPT} \citep{Dakin22}. 

The main data product are produced on the fly during the simulation and is a continuous full-sky particle light cone that extends to $z=3$. The 3D particle light-cone data are used to identify roughly 150 billion dark matter haloes using \texttt{Rockstar} \citep{Behroozi:13}, and to create all-sky 2D maps of dark matter in 200 tomographic redshift shells between $z=0$ and $z=99$. The halo catalogue and the set of 2D dark matter maps are the main inputs for the Flagship mock galaxy catalogue. For a detailed description of the production of the catalogue, see \cite{EuclidSkyFlagship}, but the main steps can be summarised as follows. First, galaxies are generated following a combination of halo occupation distribution (HOD) and abundance matching techniques. Following the HOD prescription, haloes are populated with central and satellite galaxies. Each halo contains a central galaxy and a number of satellites that depends on the halo mass.  The halo occupation is chosen to reproduce observational constraints of galaxy clustering in the local Universe \citep{Zehavi2011}. 

The luminosities of the central galaxies are assigned by performing abundance matching between the halo mass function of the simulation halo catalogue and the galaxy luminosity function (LF). The satellite luminosities are assigned assuming a universal Schechter LF in which the characteristic luminosity depends on the central luminosity in a way that ensures that the global LF agrees with observations. Galaxies are divided into three colour types, namely red, green, and blue, and the central and satellite galaxies in each group were distributed to match the clustering as a function of colour observed by \citet{Zehavi2011}. The radial positions of the satellites within a given halo follow a best-fit ellipsoidal Navarro--Frenk--White (NFW) profile \citep{NFW1997}. Finally, galaxy lensing properties are assigned within the Born approximation following the `onion universe' approach presented in \citet{Fosalba:08} and \citet{Fosalba:2013mra}. 

\subsubsection{Selection of spectroscopic sample}
We selected the spectroscopic catalogue from the Flagship simulation, version 2.1.10, in the following way.
Objects were selected based on their H$\alpha$ flux, with threshold $F_{\mathrm{H}\alpha,\,\mathrm{lim}} = 2\times 10^{-16}\,{\rm erg}\,{\rm cm}^{-2}\,{\rm s}^{-1}$. 
We considered two samples: a sample that only includes central galaxies and a sample with both central and satellite galaxies. 
The simulated data set covers a redshift range between $z_\text{min} = 0.9$ and $z_\text{max} = 1.8$, and we split it into four redshift bins,
centred at $\bar{z} = 1,\,1.2,\,1.4,\,1.65$, with half-bin width $\sigma_{\bar{z}} = 0.1,\,0.1,\,0.1,\,0.15$. 
This is the binning that is used for the standard analysis of the spectroscopic sample of \Euclid~\citep{EuclidSkyOverview}. 
The properties of the objects within each of these redshift bins are summarised in~\cref{tab:full-cat-prop}. Galaxy bias and local count slope are estimated as described in~\cref{sec:bias-est}.

\begin{table}
\caption{
Properties of the full catalogue. 
}
\label{tab:full-cat-prop}
\begin{center}
\begin{tabular}{c c c c}
\hline
$\bar{z} \pm \sigma_{\bar{z}}$  & $N_\mathrm{gal}/10^6$ & $b$  & $s$ \\
\hline
 \multicolumn{4}{c}{only centrals} \\
 \hline
 $1.0 \pm 0.1$  & $3.2$ & $1.51 \pm 0.02 $ &  $0.75$\\
$1.2 \pm 0.1$  & $2.4 $ & $1.73 \pm 0.03 $ & $0.84$\\
$1.4 \pm 0.1$  & $1.8$&  $2.09 \pm 0.03 $  &  $0.89$ \\
$1.65 \pm 0.15$ & $1.7$ &  $2.49 \pm 0.04 $ & $0.96$\\
\hline
 \multicolumn{4}{c}{centrals + satellites} \\
  \hline
$1.0 \pm 0.1$  & $3.6$  & $1.645 \pm 0.025 $ &  $0.79$\\
$1.2 \pm 0.1$  & $2.8$  & $1.89 \pm 0.03 $ & $0.88$\\
$1.4 \pm 0.1$  & $2.0$ &  $2.24 \pm 0.04 $  &  $0.92$ \\
$1.65 \pm 0.15$ & $2.0$ &  $2.70 \pm 0.04 $ & $0.99$\\
 \hline
\end{tabular}
\end{center}
\tablefoot{
The objects have been selected using the redshift provided by the Flagship data, which includes Doppler effects but neglects the gravitational redshift. Fluxes are not magnified, although flux magnification has a very small impact on this estimate.  The estimates of bias and local count slope are described in~\cref{sec:bias-est}. 
}
\end{table}

The fraction of satellite galaxies in the catalogue ranges between $10\%$ and $15\%$ in the four
redshift bins. Satellite galaxies are found preferentially in the proximity of the most massive haloes, and this results in a large galaxy bias when satellites are included, compared to the sample that only considers central galaxies. The values of the local count slope are only mildly affected by the presence of satellites. 

\subsection{Implementation of gravitational redshift in Flagship}
\label{sec:imple-grav}

In FS, the observed redshift is estimated from the peculiar velocities of galaxies. 
Other relativistic corrections, including gravitational redshift and the integrated Sachs--Wolfe effect, are neglected.
In this work, to include the gravitational redshift $z_{\Psi}=-\Psi/c^2$, we simply corrected the
observed redshift $z_\mathrm{obs, FS}$ provided in
the Flagship catalogue by
\begin{equation}
z_\mathrm{obs} = z_\mathrm{obs, FS}-(1 + \bar{z})\,\Psi/c^2\,,
\end{equation}
where $\bar{z}$ is the cosmological redshift, and we have neglected subdominant contributions.

To estimate the gravitational redshift we require the potential $\Psi$ at each galaxy. Since
on small scales the relativistic dipole is primarily sourced by gravitational redshift~\citep{Breton:2018wzk}, it is important that the potential is accurately modelled on these scales. 
But given that high-resolution gravitational potential maps are not provided by the Flagship
simulation, we adopted a semi-analytical approach and assumed that each halo where a galaxy is found can be modelled using a NFW density profile. 
The gravitational potential of the halo can then be estimated from the Poisson equation. The NFW gravitational potential $\Psi\equiv \Psi_\mathrm{NFW}$ at a comoving distance $R$
from the centre of the halo is 
\begin{equation}
\Psi_\mathrm{NFW}(R) = - 4\pi G (1+z) \rho_{\rm s} R_{\rm s}^2 \frac{\mathrm{ln}(1+R/R_{\rm s})}{R/R_{\rm s}}\,.
\label{eq:phiNFW}
\end{equation}
Here, $\rho_{\rm s}$ and $R_{\rm s}$ are the scale density and the scale radius, the two parameters that describe a NFW density profile. They depend on the concentration parameter, the virial radius, and the virial overdensity of the halo, see~\cite{MBW} and~\cite{Saga:2021jrh} for the exact expression of these quantities. 
For satellite galaxies, at a comoving distance $R$ from the centre of the parent halo, we estimated the 
gravitational potential from Eq.~\eqref{eq:phiNFW}. For the central galaxies, we took the limit
\begin{equation}
\Psi_\mathrm{NFW}(R \rightarrow 0) = - 4\pi G (1+z) \rho_s R_{\rm s}^2\,. \label{eq:phi-nfw-central}
\end{equation}
This phenomenological model
has been tested to be adequate for modelling the dipole~\citep{Saga:2020tqb, Saga:2021jrh}.

\subsection{Flux magnification}

The measured fluxes of galaxies depend on the intrinsic luminosity of the
object $L$ and its luminosity distance $D_\mathrm{L}$ by 
\begin{equation}
F = \frac{L}{4\pi D_\mathrm{L}^2}\,.
\end{equation}
Since the luminosity distance is affected by perturbations~\citep{Bonvin:2005ps, Challinor:2011bk}, the measured fluxes are also affected by relativistic effects such as magnification and Doppler corrections.  
While the observed fluxes in FS include observational effects such as dust extinction, magnification due to lensing and Doppler effects are not included. We corrected the fluxes to account for these effects using the relation
\begin{equation}
F_\mathrm{magn} = F_\mathrm{FS} \,\,\frac{(1+z)^4}{(1+z_\mathrm{obs})^4}\,\,\mu_{\rm L}\,,
\end{equation}
where $F_\mathrm{FS}$ are the fluxes stored in the mock, $F_\mathrm{magn}$ are the magnified fluxes, and $\mu_{\rm L}$ is the magnification factor due to gravitational lensing. This quantity can be expressed in terms of the convergence $\kappa$ and shear $\gamma = \gamma_1 + \mathrm{i}\,\gamma_2$:
\begin{equation}
\mu_{\rm L} = \frac{1}{(1-\kappa)^2 - \gamma_1^2 - \gamma_2^2}\,.
\end{equation}
The fluxes $F_\mathrm{magn}$ are therefore affected by gravitational lensing through $\mu_{\rm L}$ and by Doppler effects and gravitational redshift through $z_\mathrm{obs}$.

\subsection{Flux reference to perform split in bright and faint}
\label{sec:split}
The spectroscopic catalogue is constructed by selecting all objects based on their H$\alpha$ flux. However, \Euclid also measures the flux of galaxies in different colour bands, in particular the \YE band, \JE band, and \HE band, see~\cite{EuclidSkyNISP}, so performing this split using the H$\alpha$ flux as reference is not necessarily the optimal choice. In order to determine which quantity is more suitable to split the catalogue in a bright and faint population, we checked the correlation between the fluxes and the mass of the parent halo.

\begin{figure}
\begin{center}
\adjustbox{max width=\columnwidth}{
\subfloat[only centrals]{
  \includegraphics[clip,width=\columnwidth]{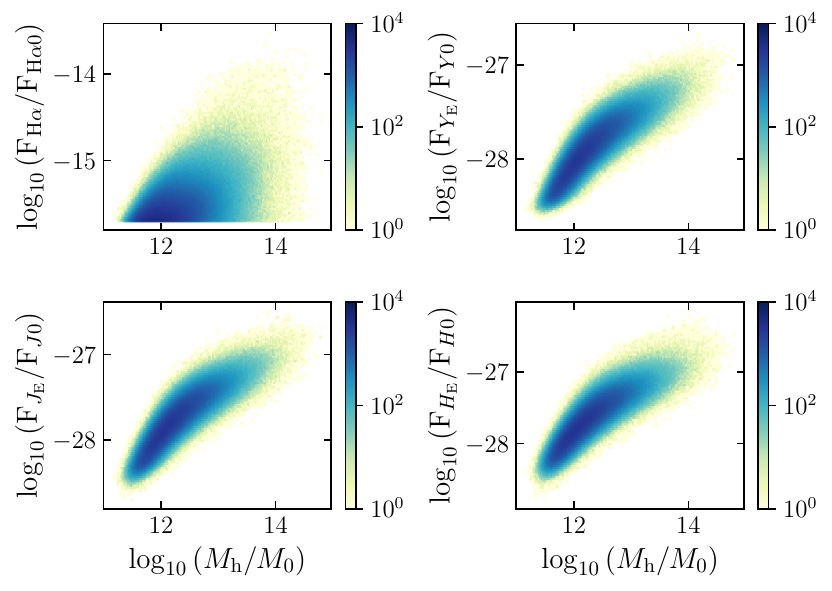}%
}
}
  \end{center}
\begin{center}
\adjustbox{max width=\columnwidth}{
\subfloat[centrals + satellites]{%
  \includegraphics[clip,width=\columnwidth]{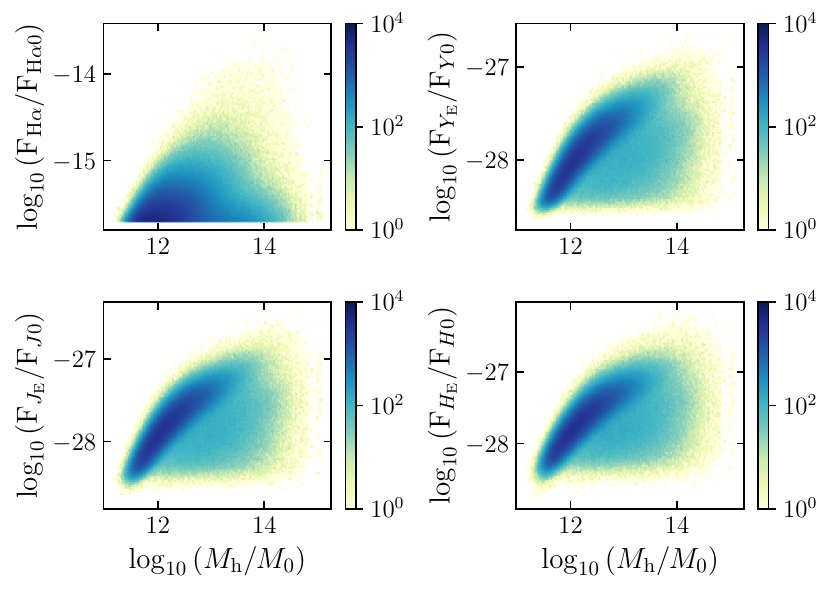}%
}
}
  \end{center}
\caption{\label{mass:corr} Correlation of galaxy fluxes with the mass of the host halo. The top panel includes only central galaxies, the bottom panel includes both central and satellite galaxies. The reference fluxes are 
$F_{\mathrm{H}\alpha 0} = 1 {\rm erg}\,{\rm cm}^{-2}\,{\rm s}^{-1}$ and $F_{Y0} = F_{J0} = F_{H0} = 1 {\rm erg}\,{\rm cm}^{-2}\,{\rm s}^{-1}\,{\rm Hz}^{-1}$,
while the reference halo mass is give by $M_0 = 1\,h^{-1}M_\odot$, where $M_\odot$ is the mass of the Sun. The colourbar highlights the density of objects in arbitrary units.}
\end{figure}

In~\cref{mass:corr} we show the correlation of different galaxy fluxes, including the H$\alpha$ flux, and the fluxes in the \YE, \JE, and \HE bands, and the mass of the host halo. 
The top panels include only central objects, while the bottom panels include both central and satellite galaxies. 
Fluxes in infrared bands are more strongly correlated with the halo mass, especially for central galaxies. This is because these fluxes are correlated with the stellar mass of the galaxy, which is correlated to the halo mass. However, the H$\alpha$ flux is correlated with the star-formation rate, which in general does not correlate with the mass of the host halo. 
For this reason, fluxes in the infrared bands are expected to be a better choice than the H$\alpha$ flux to perform the split into bright and faint populations because they better trace the mass of the host halo. Bright objects in the infrared bands are more massive than the analogue selection applied to  the H$\alpha$ flux. Therefore, the former selection leads to larger differences in galaxy bias between faint and bright objects, consequently boosting some of the contributions to the dipole.

In order to ensure that bright and faint populations have the same redshift evolution
across all bins, we did not use a single flux cut. In fact, flux is inversely proportional to the square of the luminosity distance, meaning that a fixed flux cut would lead to more bright objects at low redshift and more faint objects at high redshift.
To avoid this asymmetry in the redshift distribution, we performed the split in $20$ sub-bins, following a similar procedure to the one described in~\cite{Bonvin:2023jjq}. 

\subsection{Estimation of biases}
\label{sec:bias-est}

We describe here the method we used to estimate the galaxy bias and local count slope for the full \Euclid spectroscopic sample, 
as well as for the bright and faint populations. In our analysis, we neglect the impact of evolution bias\footnote{Our galaxy selection leads to the same observed redshift evolution of the two populations and partially cancels out the dipole contribution of source count evolution, which is proportional to the difference in the evolution biases of the two populations. Neglecting evolution bias is also fully justified by the outcome of our analysis in Sect.~\ref{sec:res-lin}, where we show that the overall linear dipole is very small.}. 

\subsubsection{Galaxy bias}
The galaxy biases for the full and the bright/faint samples is estimated using the method described in~\cite{Lepori:2022hke} and~\cite{Bonvin:2023jjq}. 
We constructed a map of the galaxy number counts using 
\texttt{HEALPix}~\citep{Gorski:2004by} and we extracted its angular power spectrum; we fitted it to a theory prediction of the power spectrum computed with the code {\sc class}~\citep{DiDio:2013bqa, DiDio:2013sea, CLASS, Blas:2011rf}, using \texttt{HMCODE} to account for nonlinear effects~\citep{Mead:2016zqy}.
The angular power spectrum of the map is estimated with the library \texttt{NaMaster}~\citep{Alonso:2018jzx}, which corrects for the effect of the mask. The \texttt{HEALPix} map is constructed from the comoving position of the objects, and therefore its angular power spectrum only includes density contributions. Thus, we fitted the galaxy bias in real space, so included only the density term in our theory model. In the fit of the simulated spectrum to the model, we fixed the cosmology so that the only free parameter is the galaxy bias. 

\subsubsection{Local count slope}

The local count slope for the entire sample is calculated from the slope of the cumulative LF of objects $N(\bar{z}, F_{\mathrm{H}\alpha} \geq F_{\mathrm{H}\alpha,\,\mathrm{lim}})$, estimated at the flux limit~\citep{Bonvin:2023jjq}.  

The impact of flux magnification is more complex for the bright and faint populations, because the two samples are constructed by applying multiple flux cuts. 
Magnification from multiple flux selections has been first discussed in~\cite{Borgeest1991}. 
In~\cref{ap:magn-bias} we derive an expression for the effective local count slope of the two populations, adapted to our case of interest,
and we test the robustness of this method. Here we report the final result. 
Since bright and faint populations are the result of two flux selections, first in
H$\alpha$ flux and, second, in \YE-band flux, we expect that for both populations,
magnification can change the count of objects in two ways: moving objects across the threshold $F_{\mathrm{H}\alpha,\,\mathrm{lim}}$ to the faint or bright population, and
transferring objects from the faint to the bright population across the threshold $F_{\YE,\,\mathrm{lim}}$. This is shown schematically in~\cref{fig:magn-flux}.
Therefore, we can define an effective count slope for the bright and faint population as the sum of two contributions,
\begin{equation}
s_{\mathrm{B/F}, \mathrm{eff}} \equiv s_{\mathrm{B/F}, \mathrm{H}\alpha} + s_{\mathrm{B/F}, \YE}\, , 
\end{equation}
where 
\begin{equation}
s_{\mathrm{B/F}, \mathrm{H}\alpha} \equiv -\frac{2}{5} \left.\frac{\partial \ln \bar{N}_\mathrm{B/F}}{\partial \ln \bar{L}_{\mathrm{H}\alpha}}\right\rvert_{{L}_{\mathrm{H}\alpha,\,\mathrm{lim}}},\quad s_{\mathrm{B/F}, \YE} \equiv -\frac{2}{5} \left.\frac{\partial \ln \bar{N}_\mathrm{B/F}}{\partial \ln \bar{L}_{\YE}}\right\rvert_{{L}_{\YE,\,\mathrm{lim}}}\,,
\end{equation}
and we introduce the LF of bright and faint galaxies in the redshift bin $\bar{z}$, that is $\bar{N}_\mathrm{B} \equiv
 N_\mathrm{B}(\bar{z},\, \bar{L}_{\mathrm{H}\alpha} \geq {L}_{\mathrm{H}\alpha,\,\mathrm{lim}},\,\bar{L}_{\YE} \geq{L}_{\YE,\,\mathrm{lim}})$
and $\bar{N}_\mathrm{F} \equiv
 N_\mathrm{F}(z,\,\bar{L}_{\mathrm{H}\alpha} \geq {L}_{\mathrm{H}\alpha,\,\mathrm{lim}},\,\bar{L}_{\YE} < {L}_{\YE,\,\mathrm{lim}})$, respectively. The luminosities $\bar{L}_{X}$ are the luminosities corresponding to the flux $F_{X}$ in the background FLRW cosmology.

Since across the flux cut $F_{Y,\,\mathrm{lim}}$ objects are transferred from the bright to the faint catalogue and vice versa, $s_{\mathrm{F}, \YE}$
 and $s_{\mathrm{B}, \YE}$ are not independent, that is,
 \begin{equation}
s_{\mathrm{F}, \YE} = - \frac{\bar{N}_\mathrm{B}}{\bar{N}_\mathrm{F}}
s_{\mathrm{B}, \YE}\,.  
\end{equation}

In practice, we estimated the slopes $s_{\mathrm{B/F}, \mathrm{H}\alpha}$ by fixing
the flux cuts in the \YE band and performing the same cuts on a version of the mock that is deeper in
$\mathrm{H}\alpha$ flux. The slope $s_{\mathrm{B}, \YE}$ is estimated after applying the selection in the
$\mathrm{H}\alpha$ flux, from the cumulative LF of the bright population at the flux cut in the
\YE band. Since we performed the split between bright and faint in sub-bins, the
\YE-band flux cut is redshift dependent. We estimated the slope of the cumulative LF in sub-bins and estimated an effective value by taking an average weighted by the number of objects in each sub-bin.

\subsection{Measurements of the dipole and its covariance}
\label{sec:meas-cov}

We used the Landy--Szalay (LS) estimator~\citep{Landy:1993yu} to measure the 3D cross-correlation between bright and faint objects within each redshift bin,

\begin{equation}
\label{eq:ls}
\xi_\mathrm{LS}(\bar{z}, d,\mu) = 
\frac{ {\rm DD}_{\rm BF} - {\rm DR}_{\rm BF} - {\rm RD}_{\rm BF} + {\rm RR}_{\rm BF}}{{\rm RR}_{\rm BF}}\,.
\end{equation}
This pair count-based estimator requires introducing two Poisson-sampled random catalogues, ${\rm R}_\mathrm{B}$ and ${\rm R}_\mathrm{F}$, that mimic the redshift distribution of the respective bright (${\rm D}_\mathrm{B}$) and faint (${\rm D}_\mathrm{F}$) data catalogues but have a uniform distribution in the solid angle. The ${\rm XY}_{\rm BF} $ in Eq.~\eqref{eq:ls} are then histograms of pair counts between catalogue ${\rm X}_\mathrm{B}$ and catalogue ${\rm Y}_\mathrm{F}$, binned in $d$ and $\mu$ and normalised by their total number of pairs, i.e., the pair counts in each bin are divided by the product of the number of objects in catalogue ${\rm X}_\mathrm{B}$ and the number of objects in catalogue ${\rm Y}_\mathrm{F}$. 

The LS estimator has been shown to have the lowest possible variance and bias among all possible estimators based on pair counts, if the correlation is small ($\xi \ll 1$) and if the number densities of the random catalogues are sufficiently high~\citep{Landy:1993yu, Kerscher:1999hc, Keihanen:2019vst}. In order to get unbiased results, we hence created random catalogues that have ten times as many objects as their corresponding data catalogues. 

In practice, we created the Poisson-sampled random catalogues from the density distribution of the data catalogues using the inversion method, as described in \cite{Schulz:2023oid}. The LS estimator is computed using a modified version of the publicly available code {\tt{CUTE}}~\citep{Alonso:2012rk}. The modifications, which are the same as in \cite{Breton:2018wzk}, extend the capabilities of {\tt{CUTE}} to allow the computation of odd multipoles in the two-point correlation function. We counted the pairs for separations $2\,h^{-1}\,\mathrm{Mpc} \leq d \leq 30\,h^{-1}\,\mathrm{Mpc}$ and orientations $-1 \leq \mu \leq 1$, binning into $d$-bins of width $\Delta d = 2\,h^{-1}\,\mathrm{Mpc}$ and $\mu$-bins of width $\Delta \mu = 0.004$. 

We did not estimate the correlation for smaller separations because such separations are not well sampled 
by our catalogues. The average separation between a bright object and the closest faint object can be derived from the cross-correlation function monopole, as it represents the average overdensity of faint objects in the neighbourhood of a bright object. From the measurements of the real-space monopoles on the light cone at small separations we found that there are typically no faint objects separated by $d \lesssim 1\,h^{-1}\,\mathrm{Mpc}$ from the average bright object, and at $\bar{z} = 1.0$, the average separation between a bright and the closest faint object is $d=3.6\,h^{-1}\,\mathrm{Mpc}$. 
The number of pairs with separation below this value is generally small, leading to very large Poisson errors 
on the measurement at those separations. Additionally, at very small separations, the LS estimator can in principle become undefined in certain ($d,\mu$)-bins if there are no pairs between the two random catalogues. For these reasons, we excluded $d < 2\,\,h^{-1}\,\mathrm{Mpc}$ from the analysis.

The $\ell$-th multipole of the LS estimator is defined by the relation
\begin{equation}
    \xi_\ell(\bar{z}, d) = \frac{2\ell+1}{2} \int_{-1}^1 \xi_\mathrm{LS}(\bar{z},d,\mu)\, L_\ell(\mu)\, \diff\mu\,.
\end{equation}
In practice, the integral is numerically computed as a sum over the $\mu$-bins. We thus computed the dipole of the LS estimator as
\begin{equation}
    \xi_\textrm{1}(\bar{z},d) \simeq \frac{3}{2}\sum_{i}^{} \xi_\mathrm{LS}(\bar{z},d,\mu_i)\,\mu_i\,\Delta \mu\,.
\end{equation}

The covariance of the measurement was estimated with the jackknife (JK) method~\citep{Norberg:2008tg},
\begin{align}
 {\rm Cov}^{\rm JK}_{ij}&:= {\rm Cov}^{\rm JK}\left(\xi_1(\bar{z}, d_i),\xi_1(\bar{z},d_j)\right)\label{eq:covJK}\\
 &= \frac{(N-1)}{N}\sum^N_{k=1}\left[\xi^k_1(\bar{z}, d_i) - \bar{\xi}_1(\bar{z},d_i)\right]\left[\xi^k_1(\bar{z},d_j) - \bar{\xi}_1(\bar{z},d_j)\right]\, ,\nonumber
\end{align}
where the survey volume at redshift $\bar{z}$ was divided into $N=50$ sub-volumes with approximately equal surface area and shape by running a {\sc kmeans}\footnote{\url{https://github.com/esheldon/kmeans_radec/}} algorithm \citep{DES:2016qvw, DES:2015eop} on one of the corresponding random catalogues. The same split was applied to the data catalogues and the LS dipole was measured $N$ times, each time removing a different sub-volume $k$ from the survey. In Eq.~\eqref{eq:covJK}, $\xi^k_1(\bar{z},d_i)$ is the dipole measured within the survey volume at redshift $\bar{z}$ from which the $k$-th sub-volume has been removed, and $\bar{\xi}_1(\bar{z}, d_i)$ is the JK sample mean,
\begin{equation}
\bar{\xi}_1(\bar{z}, d_i) = \sum^N_{k=1} \frac{\xi^k_1(\bar{z}, d_i)}{N}\, .
\end{equation}

We confirmed that increasing the number of sub-volumes from 50 to 100 does not change the estimated covariance. We conclude that with 50 JK regions, the JK covariance is already well converged and is thus a robust estimate of the measurement covariance.
\begin{figure}[t!] 
    \centering
    \includegraphics[width=\columnwidth]{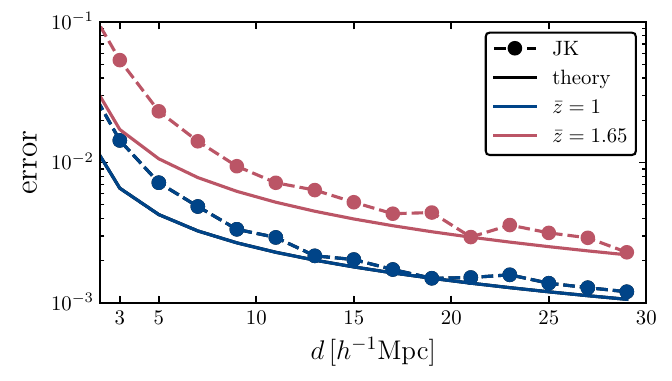}
    \caption{Comparison of the uncertainty 
    estimate from the theory covariance (continuous lines) and numerical covariance, computed from the mock using the JK resampling method (data points). We consider the two representative redshift bins centred at $\bar{z} = 1$ (in blue) and $\bar{z} = 1.65$ (in red). 
    Both estimates use a split of $90\%$ faint and $10\%$ bright.}
    \label{fig:cov-comp}
\end{figure}

In~\cref{fig:cov-comp} we compare the uncertainty 
estimate from the JK and the theory covariance. 
The theory covariance underestimates the uncertainty on all scales. In the redshift bin centred at $\bar{z} = 1$, the discrepancy between the JK and theory covariance ranges between $10\%$ and $20\%$ on scales larger than $10\,h^{-1}\,{\rm Mpc}$, while on smaller scales, the uncertainty estimated from our mock is up to a factor of two larger than the theory uncertainty. This difference is even larger at higher redshift. 
In the nonlinear regime, the  JK covariance provides a significantly more conservative forecast for the detection significance. 

\subsection{Method to estimate the significance of the detection}
\label{sec:signif}
Given a set of observations for the dipole and its theoretical prediction, we want to provide a quantitative estimate of its detection. To do this, we determined the detection in units of $\sigma$ as the difference in $\chi^2$ between accepting or rejecting the presence of the dipole. In practice, we considered the following $\chi^2$ when we include a dipole in our prediction,
\begin{equation}
    \chi^2_{\rm dipole}=\sum_{ij}\left(\xi_{1,i}^{\rm obs}-\xi_{1,j}^{\rm th}\right)^{\rm T}\left(\text{Cov}^{-1}\right)_{ij}\left(\xi_{1,i}^{\rm obs}-\xi_{1,j}^{\rm th}\right)\,,
\end{equation}
where $i$ and $j$ run over different separation bins, $\xi_{1}^{\rm th}$ stands for the theoretical prediction of the dipole, $\xi_{1}^{\rm obs}$ represents its estimate from observed data, and cov corresponds to the covariance, which can be the theoretical covariance, the JK covariance, or their combination.

We further considered the $\chi^2$ when no dipole is added into the prediction,
\begin{equation}
    \chi^2_{\rm no\, dipole}=\sum_{ij}\left(\xi_{1,i}^{\rm obs}-0\right)^{\rm T}\left(\text{Cov}^{-1}\right)_{ij}\left(\xi_{1,i}^{\rm obs}-0\right)\,.
\end{equation}
The detection significance in units of $\sigma$ can then be expressed as
\begin{equation}
    \text{detection}\,[\sigma]=\sqrt{\Delta \chi^2}=\sqrt{\chi^2_{\rm no\,dipole}-\chi^2_{\rm dipole}}\,.
\end{equation}

It is important to note that the computation of the detection significance requires the inversion of the covariance matrix. However, JK covariances may contain a non-negligible level of noise, which can lead to biased results. We corrected for this by applying the Hartlap factor~\citep{Hartlap}. In more detail, we multiplied the inverse covariance matrix by $(N_{\rm JK}-p-2)/(N_{\rm JK}-1)$, where $N_{\rm JK}$ stands for the number of JK regions and $p$ represents the number of separation bins.

\section{Results}
\label{sec:4-results}
In this section, we report the main results of the paper. 

In~\cref{sec:res-lin}
we discuss the signal-to-noise ratio ($\snr$) of the relativistic dipole in the linear regime. In~\cref{sec:res-nl}
we present the measurement of the dipole from the Flagship simulation on small scales, for separations smaller than $30\,h^{-1}\,\mathrm{Mpc}$. We consider two ways of splitting the catalogue of \Euclid into a bright and faint population, and we report the significance of detection for these two cases. 
In~\cref{sec:res-cont} we isolate the different contributions to the dipole and discuss the contribution of nonlinear velocities. 
In~\cref{sec:res-sat-magn} we investigate the impact of satellite objects and flux magnification on the significance of detection. 

In~\cref{tab:spec} we report the details for two representative splits
considered in this work: a case with $90\%$ of faint objects and $10\%$ of bright objects, and a case with $50\%$ of bright and faint objects. 
The table shows the number of objects in each redshift bin, the galaxy bias and local count slope estimates, and the mean gravitational potential at the position of the galaxy. These quantities have been used to compute the theory model in the linear and nonlinear regime and to estimate the theory covariance. 
We report the measurements for our selection of objects including only central galaxies, as well as both central and satellite galaxies.

\subsection{Relativistic dipole in the linear regime}
\label{sec:res-lin}

\begin{figure}[t!] 
    \centering
    \includegraphics[width=\columnwidth]{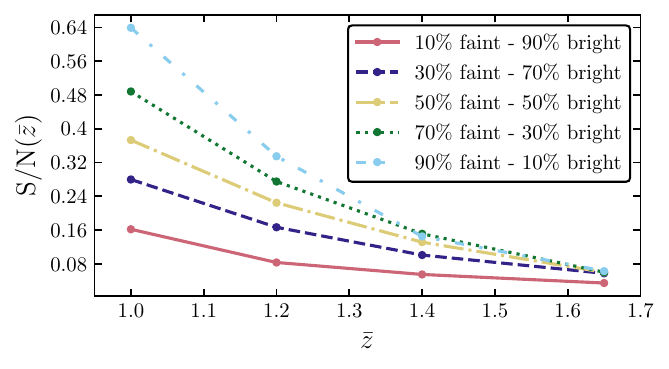}
    \caption{S/N of the dipole on linear scales ($\geq20\,h^{-1}\,\mathrm{Mpc}$) for a range of population splits. Here the model prediction is given by linear theory.}
    \label{fig:sn:linear}
\end{figure}

As a preliminary study, we examine the expected S/N of the dipole for the spectroscopic catalogue of \Euclid on large scales.
The S/N is estimated in each redshift bin as
\begin{equation}
\snr\,(\zmean_i)\,
:=
\sqrt{
\sum_{j, k}
\xi_{1}(\zmean_i, d_j)\,\,
{\rm Cov}^{-1}\left[\xi^j_{1}, \xi^k_{1}\right](\zmean_i)\,\,
\xi_{1}(\zmean_i, d_k)
}\,, \label{eq:snlin}
\end{equation}
where the dipole $\xi_{1}(\zmean, d_j)$ and  covariance ${\rm Cov}\left[\xi^j_{1}, \xi^k_{1}\right]$, given by Eq.~\eqref{eq:cov-theory},
are both computed using linear theory.
We evaluate Eq.~\eqref{eq:snlin} including separations between $d_{\rm min} \equiv 20\,h^{-1}\,\mathrm{Mpc}$ and $d_{\rm max} \equiv 200\,h^{-1}\,\mathrm{Mpc}$, and a bin width $\Delta d =
2\,h^{-1}\,\mathrm{Mpc}$.
We use the sky coverage of the Flagship simulation ($f_{\rm sky} = 0.125$).

In~\cref{fig:sn:linear} we show the $\snr$ as a function of the mean redshift, for a range of splits with different percentages of bright and faint objects. 
The $\snr$ increases as we increase the percentage of faint objects; these configurations maximise the differences between biases.
We find that the $\snr$  is larger in the low-redshift bins. However, for all configurations and redshifts, we find a $\snr < 1$, which indicates that we cannot detect the dipole in the catalogue of \Euclid in the linear regime.
This is not so surprising: the linear model of the dipole contains a Doppler term $\sim c/(r\HH)$ that boosts the signal-to-noise ratio at low redshift but decays at high redshift. 
The forecast shown in~\cref{fig:sn:linear} includes only central objects. We checked that including satellite galaxies does not change this picture substantially. We also verified that measurements of the dipole from the Flagship simulation on large scales are consistent with zero within their error bars. 
Since the relativistic dipole is not detectable by \Euclid on large scales, the rest of this analysis focuses on the nonlinear regime.

\subsection{Relativistic dipole in the nonlinear regime}
\label{sec:res-nl}
In this section, we present the measurements of the relativistic dipole on small scales from the Flagship simulation. 
We split the spectroscopic mock catalogue into two populations based on the $Y$-band flux of the objects, as described in~\cref{sec:split}.  
We consider two representative cases: a split with equal numbers of bright and faint objects ($50\%$-$50\%$ case)
and a split with $10\%$ of bright objects and $90\%$ of faint objects ($10\%$-$90\%$ case).

\begin{figure*}[t!] 
    \centering
    \includegraphics[width=\textwidth]{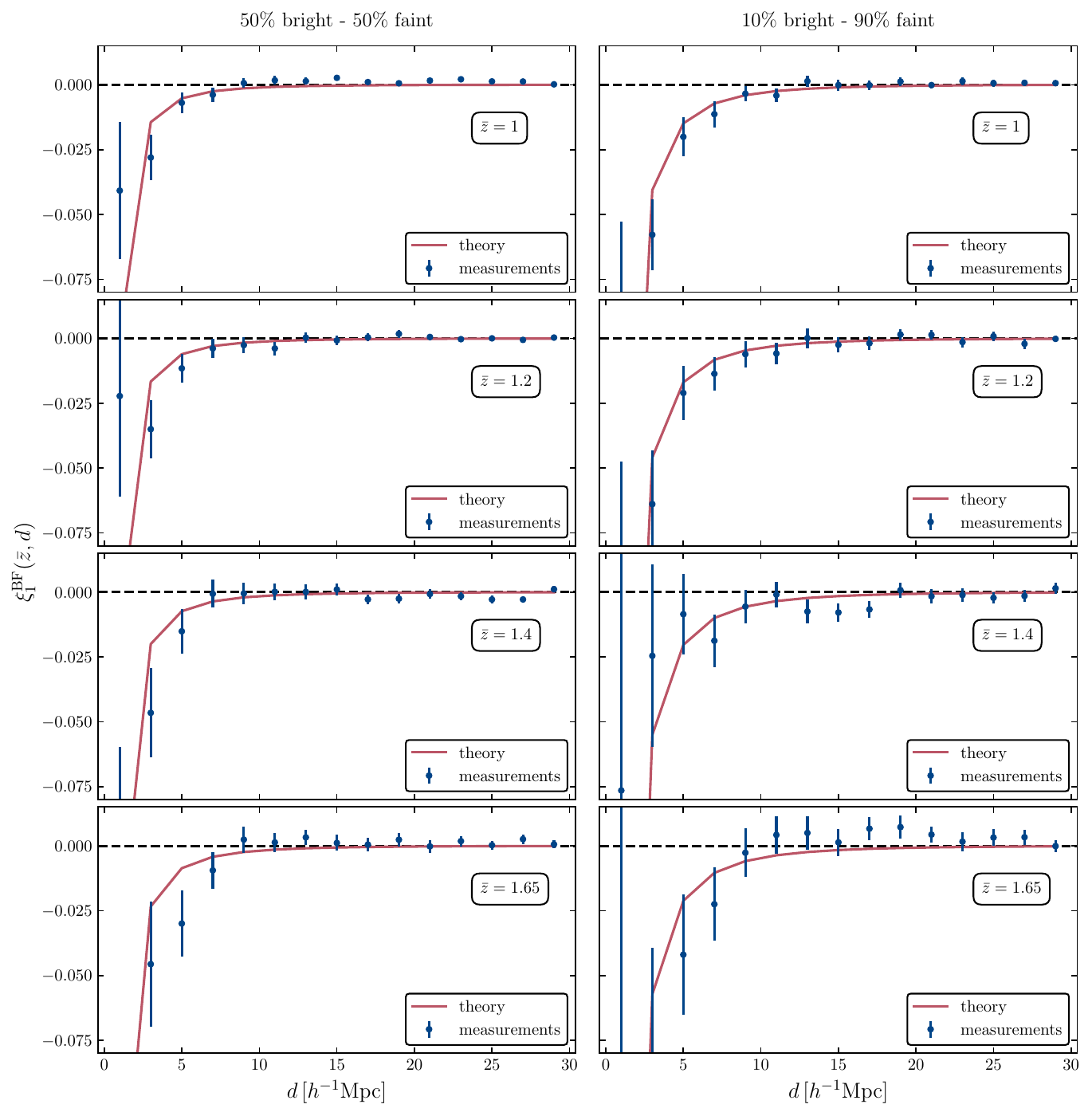}
    \caption{Measurements of the dipole on small scales, compared to the phenomenological theory model described in~\cref{sec:model-nl}. The left panels refer to the $50\%$-$50\%$ split, the right panels refer to the $10\%$-$90\%$ split. The rows from top to bottom correspond to the measurements at redshift $\bar{z} = 1, 1.2, 1.4, 1.65$, respectively. 
    }
    \label{fig:meas-splits}
\end{figure*}

In~\cref{fig:meas-splits} we show the measurement of the dipole in the four redshift bins at $\bar{z} = 1, 1.2, 1.4, 1.65$, for the $50\%$-$50\%$ case (left panels) and $10\%$-$90\%$ case (right panels). 
The catalogue processed to obtain these measurements only contains central galaxies, and we neglect flux magnification. 
The predictions of theory is estimated from the model in~\cref{sec:model-nl}, which assumes that the kinematic contributions to the dipole are adequately modelled using linear theory, with nonlinearities mainly affecting the gravitational redshift contribution. 
The specifics of the galaxy populations (galaxy bias and mean gravitational potential at the galaxy position) are estimated from the catalogues and their values are reported in~\cref{tab:spec}. 

The data points agree with the prediction of the theory roughly within two standard deviations on scales $d \geq 3\,h^{-1}\,\mathrm{Mpc}$ for both cases. The fact that the first data point at $d = 1\,h^{-1}\,\mathrm{Mpc}$ is rather off compared to the theory prediction is not that surprising, as very few pairs of bright and faint objects are found at this separation.

\begin{table*}[ht]
  \centering
  \caption{\label{tab:det-sign}
 Detection significance. 
  }
\begin{tabular}{ccccccc}
    \hline
    & \multicolumn{2}{c}{theory covariance} & \multicolumn{2}{c}{JK covariance} & \multicolumn{2}{c}{combined covariance}\\
    Redshift & $50\%$ - $50\%$ & $10\%$ - $90\%$  & $50\%$ - $50\%$ & $10\%$ - $90\%$   & $50\%$ - $50\%$ & $10\%$ - $90\%$  \\
    \hline
    $1$      & $6.5\sigma$ & $9.2\sigma$  &$3\sigma$ & $4\sigma$   & $3.1\sigma$ & $4.5\sigma$   \\
    $1.2$    &  $5.8\sigma$  & $7.3\sigma$   & $2.9\sigma$ & $3.4\sigma$  & $3.2\sigma$ & $3.7\sigma$    \\
    $1.4$    & $5\sigma$  & $1.6\sigma$   & $2.3\sigma$ & No detection  & $2.5\sigma$ & $1.5\sigma$    \\
    $1.65$   & $4.9\sigma$ & $6\sigma$   & $2.2\sigma$& $ 2.1\sigma$   & $2.2\sigma$ & $2.1\sigma$    \\
    \hline
    Total    & $11.2 \sigma$ & $13.3 \sigma$ &  $5.2 \sigma$ & $5.7 \sigma$ &  $5.6 \sigma$   &  $6.4\sigma$  \\
     \hline
\end{tabular}
\tablefoot{ We report the detection significance in each redshift bin, for the $50\%$-$50\%$ split and the $10\%$-$90\%$ split. The columns labelled with `theory covariance' use the linear covariance described in~\cref{sec:cov}. The columns labelled with  `JK covariance' use the numerical covariance estimated with the JK method, as detailed in~\cref{sec:meas-cov}. 
The `combined covariance' corresponds to the rescaling of the theory covariance matrix by the diagonal elements of the JK covariance.
 }
\end{table*}

As expected, in the $50\%$-$50\%$ case, we find that the amplitude of the signal is smaller compared to the $10\%$-$90\%$ case. However, the errors in the measurements are larger in the $10\%$-$90\%$ case due to larger shot noise in the reduced bright sample. 
To assess which split gives an overall better S/N, we follow the method described in~\cref{sec:signif} to quantify the significance of detection in the two cases. 
In~\cref{tab:det-sign} we report the values of the detection significance for three possible choices for the covariance: the theory covariance, the JK covariance, and a combination of the two. The combined covariance is constructed as~\citep[see, e.g.,][]{Bonvin:2023jjq}
\begin{equation}
{\rm Cov}^{\rm comb}_{ij} = {\rm Corr}^{\rm th}_{ij} \sqrt{ {\rm Cov}^{\rm JK}_{ii}{\rm Cov}^{\rm JK}_{jj}}\, ,
\end{equation}
where ${\rm Corr}^{\rm th}_{ij}$ is the theory correlation matrix, given by
\begin{equation}
{\rm Corr}^{\rm th}_{ij} \equiv {\rm Cov}^{\rm th}_{ij} \:/ \sqrt{ {\rm Cov}^{\rm th}_{ii}{\rm Cov}^{\rm th}_{jj}}\, ,
\end{equation}
and $i$ and $j$ run over the components of the covariance matrix. 
The combined covariance provides the same error estimate as the JK method, having the advantage that the non-diagonal elements do not exhibit spurious numerical oscillations and follow the trend expected from the theory covariance. 
Assuming the measurements in different redshift bins are independent, we also computed the total detection significance by summing in quadrature the significance of all bins. The first data point at $d = 1\,h^{-1}\,\mathrm{Mpc}$ is excluded from this analysis. 

The JK covariance and the combined covariance give roughly consistent results, forecasting a detection significance between $3\,\sigma$ and $4\,\sigma$ in the redshift bins $\bar{z} = 1, 1.2$ and a detection significance $\lesssim 2\,\sigma$ at higher redshift. 
The total detection significance amounts to 5\,$\sigma$--6\,$\sigma$. 
Overall, the $10\%$-$90\%$ split leads to slightly better detection significance. Therefore, in the remainder of this paper, we consider this case our baseline. 
The theory covariance gives a considerably better detection significance, ranging between $11\,\sigma$ ($50\%$-$50\%$ split) and $13\,\sigma$ ($10\%$-$90\%$ split), when all the redshift bins are combined.   
As a conservative choice, we consider the combined covariance as the fiducial. 

The estimation of the detection significance assumes perfect knowledge of the linear galaxy bias of the two populations. While we can measure this quantity very accurately in simulations, where the underlying cosmological model is known, this will not be possible with the \Euclid real data. This issue can be overcome by considering a joint analysis of the even multipoles of both galaxy populations combined with the dipole measurement, leaving the galaxy bias as free parameters in the analysis.
It is shown in~\cite{Sobral-Blanco:2022oel} that the bias of the bright and faint populations are typically very well constrained by the even multipoles and that the uncertainty on the biases barely impacts the measurement of gravitational redshift.
More precisely, comparing the detectability of gravitational redshift with the detectability of gravitational redshift multiplied by the bias difference $b_{\rm B}-b_{\rm F}$ leads to a difference in precision of 0.1\%. This shows that even if the bias difference is fully degenerated with the gravitational redshift signal in the dipole, the biases are so well constrained by the even multipoles that the degeneracy can be extremely well broken.

\subsection{Isolating the different contributions to the dipole}
\label{sec:res-cont}

In this section, we study the different physical effects that contribute to the nonlinear dipole. 
We adopt the baseline split ($10\%$-$90\%$). 
In order to isolate Doppler effects and gravitational redshift, we measure the dipole from the Flagship catalogue including different contributions to the observed redshift. We consider the following combinations: a) the only contribution to the redshift comes from the Hubble expansion (background redshift), b) we include only the effect of peculiar velocities (and neglect gravitational redshift), c) we include only the effect of gravitational redshift, and d) we include both peculiar velocities and gravitational redshift. 

Note that case a) does not completely remove the effect of peculiar velocities on the dipole due to the so-called \textit{light-cone} effect~\citep{Kaiser:2013ipa,Bonvin:2013ogt, Breton:2018wzk}. This arises because the rate (in terms of look-back time, for example) at which a light ray intercepts sources depends on the peculiar motion of the sources, and it is higher when the sources move towards the light ray (away from the observer). This information is carried back to the observer, who consequently sees an enhanced (suppressed) number density in regions where sources are moving away from (towards) the observer. For example, on the near side of a cluster where infalling galaxies move away from us, the density is seen to be higher, whereas on the far side it is lower. The overall effect is a dipole (which is opposite in sign to the one from the gravitational redshift).
Thus, all a), b), c), and d) are affected by the light-cone effect but only b) and d) explicitly include the Doppler contribution
to the redshift. 

\begin{figure*}[t!] 
    \centering
    \includegraphics[width=\textwidth]{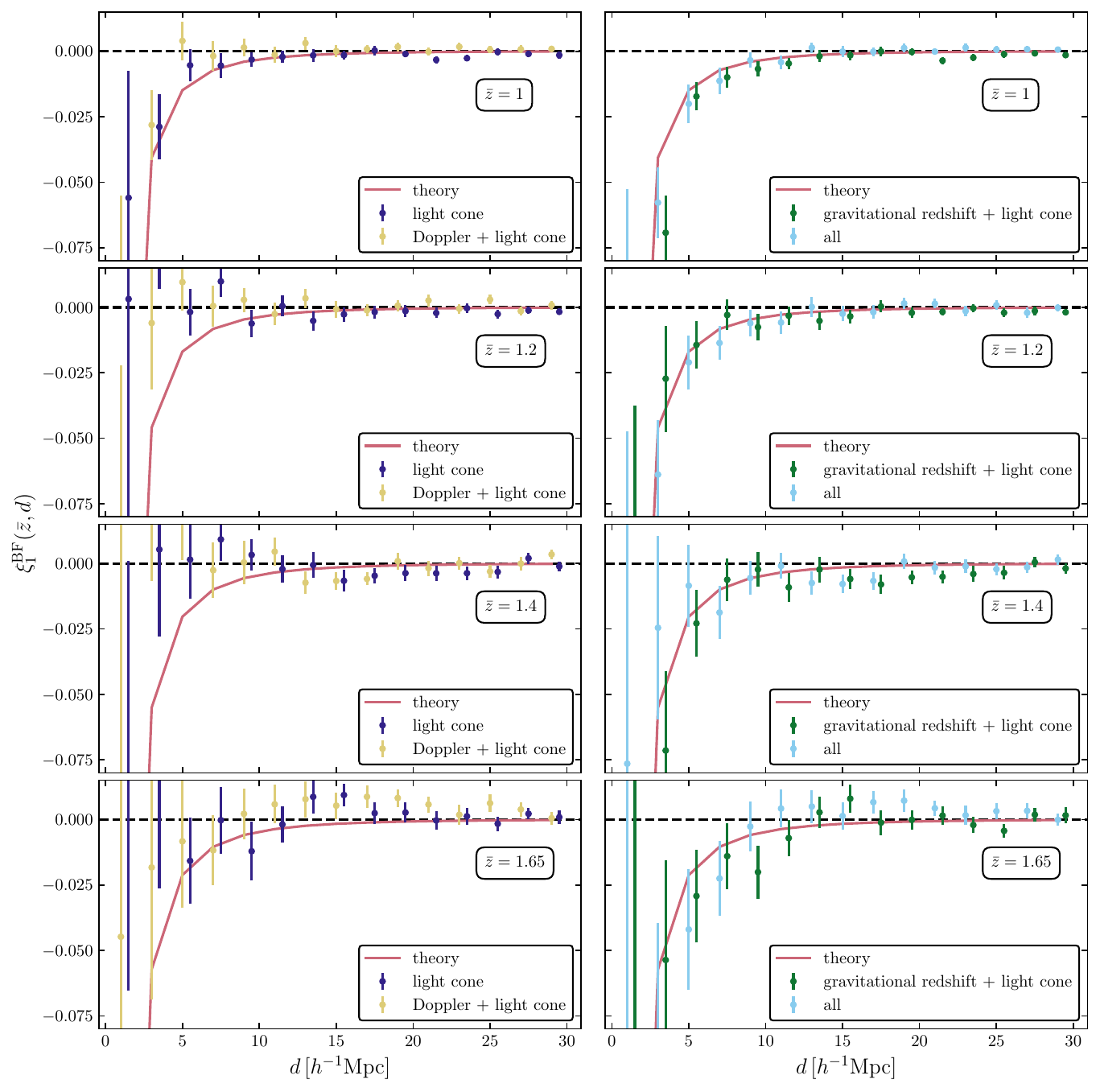}
    \caption{Measurements of the dipole isolating different contributions, using the $10\%$ bright - $90\%$ faint split. In the left panels, we show the `light cone' data points from the catalogue using the background redshift to perform both selection and measurements, while `Doppler + light cone' uses the redshift corrected by the peculiar velocities of the galaxies. In the right panels, `gravitational redshift + light cone' uses a redshift that includes gravitational redshift but does not account for the peculiar motion of the sources, and  the measurements labelled as `all' include both peculiar velocities and gravitational redshift.  The theory predictions (red continuous lines) represent our fiducial theory model, which incorporates linear velocities and nonlinear gravitational redshift. Rows from top to bottom correspond to measurements at redshift $\bar{z} = 1, 1.2, 1.4, 1.65$, respectively. The `light cone' and `gravitational redshift + light cone' data points have been offset by $0.5\,h^{-1}\,{\rm Mpc}$ along the $d$-axis for better visualisation.}
    \label{fig:meas-cont}
\end{figure*}

In~\cref{fig:meas-cont} we show the measurements for a) and b) in the left panels, and for c) and d) in the right panels, compared to our theory prediction that includes linear velocities and nonlinear gravitational potential. 
The two contributions of peculiar velocities, coming from the light-cone and Doppler effects, are individually consistent with zero at $\bar{z} = 1.2, 1.4, 1.65$.  
In the redshift bin $\bar{z} = 1$, the light-cone effect is not completely negligible, as the dipole measurements exhibit a systematic negative amplitude on scales below $10\,h^{-1}\,\mathrm{Mpc}$. However, the combined Doppler and light-cone effects lead to a dipole consistent with zero on scales $> 3\,h^{-1}\,\mathrm{Mpc}$. 
This suggests that at $\bar{z} = 1$ the contributions from nonlinear Doppler effects and
the light-cone effects are individually not negligible. However, since they generate a dipole of opposite sign, their combined contribution is consistent with zero on scales $> 3\,h^{-1}\,\mathrm{Mpc}$. 
Since both of these effects are coupled to the gravitational redshift
in the full dipole measurements, we cannot robustly conclude that
the effect of nonlinear peculiar velocities is negligible at $\bar{z} = 1$, where we forecast the greater detection significance. A model that can consistently treat both nonlinear velocities and nonlinear gravitational potential is needed for a correct interpretation of the dipole measurement at these scales. 

\subsection{Impact of satellite galaxies and flux magnification}
\label{sec:res-sat-magn}

In our baseline settings, we have only included central galaxies, and we have neglected
flux magnification. In this section, we discuss the impact of satellite galaxies and flux magnification on the dipole measurements, which includes both the Doppler and the subdominant contribution from lensing magnification. 

\begin{figure*}[t!] 
    \centering
    \includegraphics[width=\textwidth]{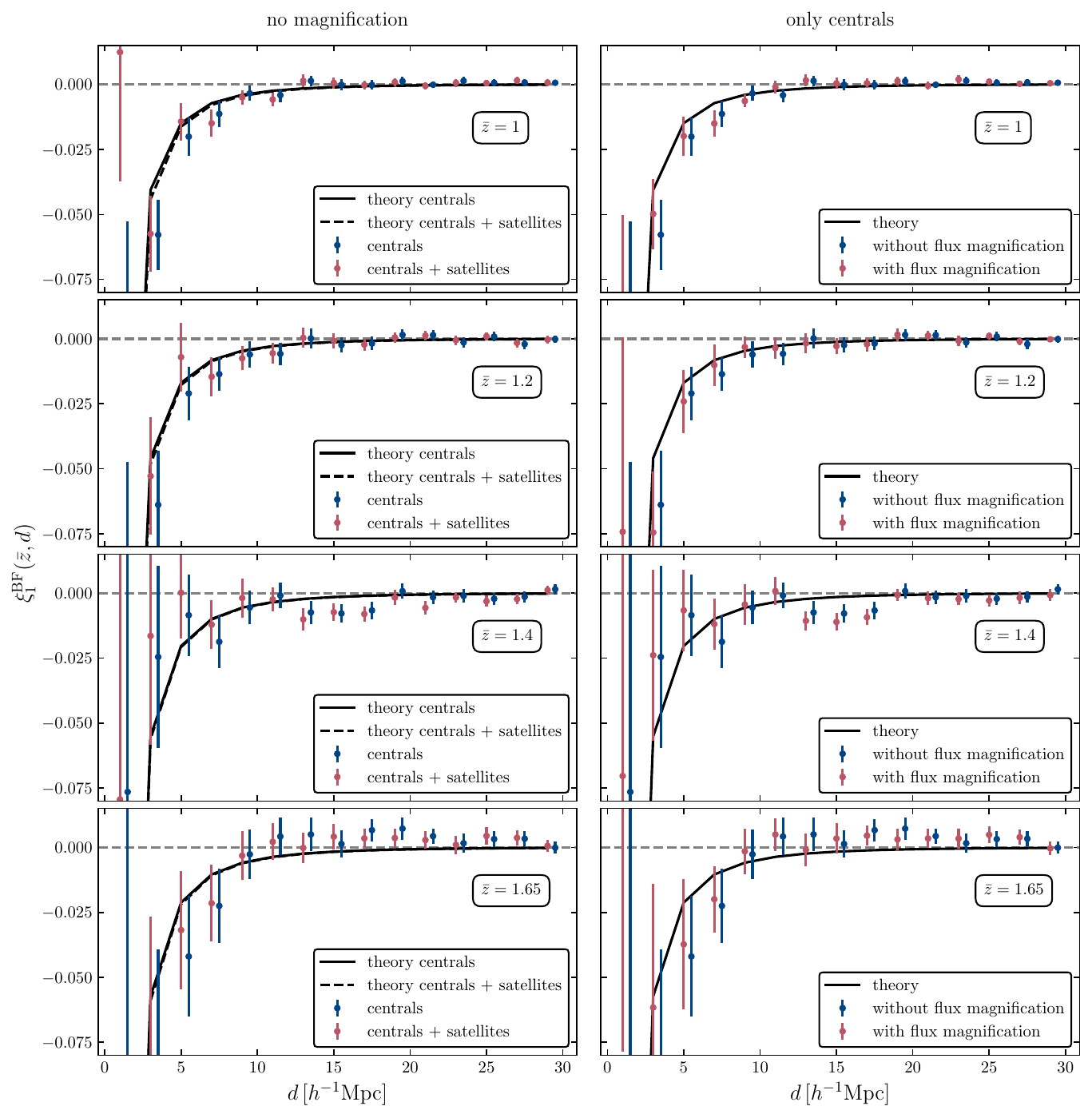}
    \caption{Impact of satellite galaxies (left panels) and flux magnification (right panels) on the measurements, for the $10\%$ bright - $90\%$ faint split. Since in our fiducial model velocities are treated linearly, flux magnification has a negligible impact on the theory prediction. Rows from top to bottom correspond to measurements at redshift $\bar{z} = 1, 1.2, 1.4, 1.65$, respectively. The `centrals' and `without flux magnification' data points have been offset by $0.5\,h^{-1}\,{\rm Mpc}$ along the $d$-axis for better visualisation.  }
    \label{fig:meas-sat-magn}
\end{figure*}

We performed the dipole measurements for our baseline split ($10\%$-$90\%$) including satellite objects in the sample (and neglecting flux magnification). The specifics for the catalogues with satellites are reported in~\cref{tab:spec}.
Compared to the selection with only central objects, including satellites results in a larger number of objects, larger galaxy bias, and higher average gravitational potential at the positions of the sources for both populations. This is because satellite galaxies are primarily found inside the most massive haloes of the simulation.

In the left panels of~\cref{fig:meas-sat-magn} we compare the dipole measurements and theory prediction for the cases with and without satellite galaxies. Both measurements and theory prediction change very little when satellite objects are included. 
In~\cref{tab:det-sign-sat-magn} we report the detection significance for measurements with satellite objects (second column). We see that the detection significance is systematically slightly decreased in all redshift bins. 
However, this does not change the outcome of the analysis substantially: we forecast a detection significance of $\approx 4\,\sigma$
and $\approx 3\,\sigma$ in the bins centred at $\bar{z} = 1$ and $\bar{z} = 1.2$, respectively. At higher redshift, the detection significance is below $2\,\sigma$.

\begin{table*}[ht]
  \centering
  \caption{\label{tab:det-sign-sat-magn}
 Impact of satellite galaxies and flux magnification on the detection significance. 
  }
\begin{tabular}{cccc}
    \hline
    redshift & \multicolumn{1}{c}{centrals, no flux magnification} & \multicolumn{1}{c}{centrals + satellites, no flux magnification} & \multicolumn{1}{c}{centrals, with flux magnification}\\
    \hline
    $1$      & $4.5\,\sigma$  & $4.4\,\sigma$ &   $4.7\,\sigma$ \\
    $1.2$    & $3.7\,\sigma$  & $2.8\,\sigma$ &   $3.4\,\sigma$ \\
    $1.4$    & $1.5\,\sigma$  & $< 1\,\sigma$ &  $1.4\,\sigma$  \\
    $1.65$   & $2.1\,\sigma$  & $1.8\,\sigma$ &    $1.8\,\sigma$   \\
    \hline
\end{tabular}
\tablefoot{
For this estimate, we have considered the $10\%$ bright - $90\%$ faint split and used the `combined' covariance.
}
\end{table*}

We perform a similar analysis to test the impact of flux magnification. 
In the right panels of~\cref{fig:meas-sat-magn} we compare the dipole measurements for the cases with and without flux magnification. Flux magnification does not affect the gravitational redshift contribution to the dipole, and since this is by far the largest contribution to our prediction, its impact on the theory model is largely negligible. The measured dipole for the catalogues with flux magnification is consistent with the measurements without flux magnification. In~\cref{tab:det-sign-sat-magn} we report the significance of the detection for these measurements.  
We find that it is compatible with the estimate for our baseline setting, with only central galaxies and no flux magnification. The differences between the two cases are $\leq 0.3\,\sigma$ in all redshift bins. 

Overall, we find that including satellites or flux magnification does not change the outcome of the analysis in a relevant way. 

\section{Conclusions}
\label{sec:conc}

In this paper, we present a measurement of the dipole of the two-point correlation function in the Flagship simulation, for a selection of objects tailored to the Euclid Spectroscopic Survey. 
This observable is sourced by gravitational redshift and Doppler effects, and it
has been shown that it can be used to test the equivalence principle on cosmological scales, see, for example,~\cite{Bonvin:2018ckp}.

Since the contribution of gravitational redshift to the observed redshift is not currently included in the Flagship mock, we implemented this effect by modelling the gravitational potential at the position of each galaxy using
the NFW profile for the host haloes.
In order to isolate these relativistic effects from the standard clustering contributions, given by the galaxy overdensity and RSDs, we split the full sample of galaxies into two populations. 
The split is performed using their fluxes in the $\YE$ band as reference. We show that this choice leads to larger differences in galaxy bias for the bright and faint sample, being more correlated with the mass of the host halo than the ${\rm H}\alpha$ flux. 
On large scales, the dipole measured from the Flagship mock is consistent with zero regardless of the
percentage of bright and faint objects. This is not surprising, as a signal-to-noise analysis based on linear theory leads to $\snr< 1$. 
On small scales, the amplitude of the signal is significantly larger due to the contributions of nonlinear gravitational redshift. Using the model presented in~\cite{Saga:2020tqb, Saga:2021jrh} as a reference, on scales $d < 30\,h^{-1}\,{\rm Mpc}$ and for our fiducial split with $90\%$ faint objects and $10\%$ bright objects, we found a detection significance of $4.5\,\sigma$, $3.7\,\sigma$, $1.5\,\sigma$, and $2.1\,\sigma$ in the four redshift bins $\bar{z} = 1, 1.2, 1.4, 1.65$, respectively. 
The overall detection significance amounts to $6\,\sigma$. 
These values are obtained from the combined covariance, which rescales the analytical covariance to match the uncertainty 
estimated from the mock using the JK method. This is our fiducial choice, which leads to a conservative estimate.
A split with equal number of bright and faint objects leads to a slightly lower detection significance at low redshift.

While our nonlinear model accounts for nonlinear gravitational redshift, it models
velocities using linear theory. We tested this assumption by separating the different contributions to the dipole: the light-cone effect, the sum of light-cone and Doppler effects, and the gravitational redshift. Although the gravitational redshift is the largest contribution to the nonlinear dipole, the light-cone effect is not completely negligible at $\bar{z} = 1$.  Nevertheless, the contributions from light-cone and Doppler effects appear to roughly cancel out. We cannot interpret this cancellation in our nonlinear model, since it only models velocities using linear theory. To correctly  
separate the different contributions to the dipole at low redshift and on small scales, we require a model that can consistently treat both the nonlinearities in the velocities and the gravitational potential~\citep[see, e.g.,][]{Dam:2023cem}. We leave the development of a complete nonlinear model to future work.

We also studied the impact of satellite galaxies and flux magnification on the estimated dipole, finding that the detection significance of the dipole is not substantially affected by them. 
Although flux magnification has negligible impact on our measurements, in~\Cref{ap:magn-bias} we validate different methods to estimate the local count slope for selections with multiple flux cuts. This work is relevant for modelling lensing magnification in galaxy clustering when the selection of galaxies is not a simple flux cut. 
The measurements in the Flagship mock catalogue cover $1/8$ of the sky, which roughly corresponds to $5000\,{\rm deg}^2$. This is about twice the sky coverage planned for Euclid DR1. Assuming that the number density and galaxy bias of the Flagship mock reflects the properties of the galaxies observed by \Euclid in all the three planned Data Releases~\citep{Scaramella-EP1, EuclidSkyOverview}, we forecast that a detection of the dipole is foreseeable for DR2 and DR3. 
In estimating the detection significance of the dipole, we have assumed that the \Euclid spectroscopic sample is pure and complete. A detailed analysis accounting for these effects and other systematic uncertainties is left for future work.

While we include in our measurements scales $d < 30\,h^{-1}\,{\rm Mpc}$, we point out that most of the signal-to-noise comes from separations $d \sim 5\,h^{-1}\,{\rm Mpc}$. This is the typical scale at which gravitational redshift has been detected in galaxy clusters; see, for example,~\cite{Wojtak_2011},~\cite{Sadeh:2014rya}, and~\cite{Rosselli:2022qoz}. The key observable in these measurements is the distribution of line-of-sight (LOS) velocities for cluster member galaxies around the cluster centre: the kinematic Doppler effect gives rise to a broad symmetric distribution around a zero LOS velocity, while gravitational redshift shifts the value of the mean velocity of this distribution.
Compared to the cluster measurements, the method we propose has the advantage of not requiring the identification of cluster member galaxies or associating the cluster centre with a specific galaxy in the spectroscopic catalogue. Measuring the galaxy cross-correlation function is, in a way, more general, as various splitting strategies can be tested and combined to optimise the measurements. The different systematics affecting the two observables make them highly complementary.
A quantitative comparison between the two approaches would require forecasting the detectability of gravitational redshift in the \Euclid cluster catalogue~\citep{Sartoris:2015aga, Adam-EP3}, which is beyond the scope of this work.

%

\begin{acknowledgements}
The work of FL, SS, JA, and CB is supported by the Swiss National Science Foundation. CB and LD acknowledge support from the European Research Council (ERC) under the European Union’s Horizon 2020 research and innovation program (grant agreement No. 863929; project title “Testing the law of gravity with novel large-scale structure observables”). 
\AckEC 

This work has made use of CosmoHub~\citep{TALLADA2020100391, 2017ehep.confE.488C}.

CosmoHub has been developed by the Port d'Informacio Cientifica (PIC), maintained through a collaboration of the Institut de Fisica d'Altes Energies (IFAE) and the Centro de Investigaciones Energeticas, Medioambientales y Tecnologicas (CIEMAT) and the Institute of Space Sciences (CSIC \& IEEC).
CosmoHub was partially funded by the "Plan Estatal de Investigacion Cientifica y Tecnica y de Innovacion" program of the Spanish government, has been supported by the call for grants for Scientific and Technical Equipment 2021 of the State Program for Knowledge Generation and Scientific and Technological Strengthening of the R+D+i System, financed by MCIN/AEI/ 10.13039/501100011033 and the EU NextGeneration/PRTR (Hadoop Cluster for the comprehensive management of massive scientific data, reference EQC2021-007479-P) and by MICIIN with funding
from European Union NextGenerationEU(PRTR-C17.I1) and by Generalitat de Catalunya.
\end{acknowledgements}

%
%

\bibliography{Euclid}

%

\begin{appendix}
\onecolumn 
  
\section{Validation of the effective local count slope estimates}
\label{ap:magn-bias}

Here we derive an expression for the effective local count slope of the bright and faint population.
We identify the number of bright objects in a redshift bin, at a given angular position in the sky, as the number of galaxies with $F_{\mathrm{H}\alpha} \geq  F_{\mathrm{H}\alpha,\,\mathrm{lim}}$ and $F_{\YE} \geq F_{\YE,\,\mathrm{lim}}$. Proceeding similarly to the case with a single flux limit, we can Taylor expand the number of objects with magnified fluxes $ N_\mathrm{B} \equiv N_\mathrm{B}(\bar{z}, F_{\mathrm{H}\alpha} \geq F_{\mathrm{H}\alpha,\,\mathrm{lim}}, F_{\YE} \geq F_{\YE,\,\mathrm{lim}})$. Denoting by $\bar{N}_\mathrm{B} \equiv
 N_\mathrm{B}(\bar{z}, \bar{L}_{\mathrm{H}\alpha} \geq {L}_{\mathrm{H}\alpha,\,\mathrm{lim}}, \bar{L}_{\YE} \geq{L}_{\YE,\,\mathrm{lim}})$\footnote{The luminosities $\bar{L}_{X}$ are the luminosities corresponding to the flux $F_{X}$ in the background.}
 the number of objects without flux magnification, 
and including only linear terms, we obtain
 \begin{equation}
 N_\mathrm{B} = \bar{N}_\mathrm{B} + \left.\frac{\partial \bar{N}_\mathrm{B}}{\partial \bar{L}_{\mathrm{H}\alpha}}\right\rvert_{{L}_{\mathrm{H}\alpha,\,\mathrm{lim}}} \delta L_{\mathrm{H}\alpha} + \left.\frac{\partial \bar{N}_\mathrm{B}}{\partial \bar{L}_{\YE}}\right\rvert_{{L}_{\YE,\,\mathrm{lim}}} \delta L_{\YE}\,.
 \end{equation}
 Using that 
 \begin{equation}
    \frac{\delta L_{\mathrm{H}\alpha}}{\bar{L}_{\mathrm{H}\alpha}} =
    \frac{\delta L_{\YE}}{\bar{L}_{\YE}} = 2   \frac{\delta D_\mathrm{L}}{D_\mathrm{L}}\,, 
 \end{equation}
where $\delta D_\mathrm{L}$ is 
the perturbation to the luminosity distance computed in~\cite{Bonvin:2005ps} and~\cite{Challinor:2011bk},  
we find that the contribution to the number counts of bright objects 
due to flux magnification is
\begin{equation}
\Delta_\mathrm{B} = \bar{\Delta}_\mathrm{B} - 5 \left(s_{\mathrm{B}, \mathrm{H}\alpha} + s_{\mathrm{B}, \YE}\right) \frac{\delta D_\mathrm{L}}{D_\mathrm{L}}\,.
\end{equation}
The parameters $s_{\mathrm{B}, \mathrm{H}\alpha}$, and $s_{\mathrm{B}, \YE}$ are defined as
\begin{equation}
s_{\mathrm{B}, \mathrm{H}\alpha} \equiv -\frac{2}{5} \left.\frac{\partial \ln \bar{N}_\mathrm{B}}{\partial \ln \bar{L}_{\mathrm{H}\alpha}}\right\rvert_{{L}_{\mathrm{H}\alpha,\,\mathrm{lim}}},
\qquad s_{\mathrm{B}, \YE} \equiv -\frac{2}{5} \left.\frac{\partial \ln \bar{N}_\mathrm{B}}{\partial \ln \bar{L}_{\YE}}\right\rvert_{{L}_{\YE,\,\mathrm{lim}}}\,.
\end{equation}
Therefore, the effective count slope from the bright population, can
be estimated as
\begin{equation}
s_{\mathrm{B}, \mathrm{eff}} \equiv s_{\mathrm{B}, \mathrm{H}\alpha} + s_{\mathrm{B}, \YE}.  
\end{equation}

Similarly, one can define an effective count slope for the faint population,
\begin{equation}
s_{\mathrm{F}, \mathrm{eff}} \equiv s_{\mathrm{F}, \mathrm{H}\alpha} + s_{\mathrm{F}, \YE},
\end{equation}
where
\begin{equation}
s_{\mathrm{F}, \mathrm{H}\alpha} \equiv -\frac{2}{5} \left.\frac{\partial \ln \bar{N}_\mathrm{F}}{\partial \ln \bar{L}_{\mathrm{H}\alpha}}\right\rvert_{{L}_{\mathrm{H}\alpha,\,\mathrm{lim}}},
\qquad s_{\mathrm{F}, \YE} \equiv -\frac{2}{5} \left.\frac{\partial \ln \bar{N}_\mathrm{F}}{\partial \ln \bar{L}_{\YE}}\right\rvert_{{L}_{\YE,\,\mathrm{lim}}}\,,
\end{equation}
and $\bar{N}_\mathrm{F} \equiv
 N_\mathrm{F}(z, \bar{L}_{\mathrm{H}\alpha} \geq {L}_{\mathrm{H}\alpha,\,\mathrm{lim}}, \bar{L}_{\YE} < {L}_{\YE,\,\mathrm{lim}})$. 
 We also note that $s_{\mathrm{F}, \YE}$
 and $s_{\mathrm{B}, \YE}$ are not 
 independent of each other, 
 because 
 magnification fluxes at the boundary $F_{\YE,\,\mathrm{lim}}$ lead 
 to a transfer of objects from the faint to the bright sample or vice versa, which implies
  \begin{equation}
\left.\frac{\partial \bar{N}_\mathrm{F}}{\partial \ln \bar{L}_{\YE}}\right\rvert_{{L}_{\YE,\,\mathrm{lim}}} = - \left.\frac{\partial \bar{N}_\mathrm{B}}{\partial \ln \bar{L}_{\YE}}\right\rvert_{{L}_{\YE,\,\mathrm{lim}}}\, ,
\end{equation}
and thus
 \begin{equation}
s_{\mathrm{F}, \YE} = - \frac{\bar{N}_\mathrm{B}}{\bar{N}_\mathrm{F}}
s_{\mathrm{B}, \YE}.  
\end{equation}

We also notice that, when the same reference
flux is used for the spectroscopic selection and for splitting the population into bright and faint
objects, that is, $F_{\YE} \rightarrow F_{\mathrm{H}\alpha}$, we have that $s_{\mathrm{B}, \mathrm{H}\alpha}$ is identically zero, and
 \begin{equation}
\left.\frac{\partial \bar{N}_\mathrm{F}}{\partial \ln \bar{L}_{\mathrm{H}\alpha}}\right\rvert_{{L}_{\mathrm{H}\alpha,\,\mathrm{lim}}} = \left.\frac{\partial \bar{N}}{\partial \ln \bar{L}_{\mathrm{H}\alpha}}\right\rvert_{{L}_{\mathrm{H}\alpha,\,\mathrm{lim}}},
\end{equation}
where $\bar{N} \equiv N (\bar{z}, \bar{L}_{\mathrm{H}\alpha} \geq {L}_{\mathrm{H}\alpha})$ is the number of objects in the redshift bin with non-magnified fluxes
above the survey limit. 
In this particular case, we also have
 \begin{equation}
s_{\mathrm{F}, \mathrm{H}\alpha}  = \frac{\bar{N}}{\bar{N}_\mathrm{F}}
s,  \qquad s_{\mathrm{F}, \mathrm{eff}} \equiv  - \frac{\bar{N}_\mathrm{B}}{\bar{N}_\mathrm{F}}
s_{\mathrm{B}, \YE} + \frac{\bar{N}}{\bar{N}_\mathrm{F}}s,
\end{equation}
where $s$ is the local count slope of the full sample. Therefore, we recover the result
derived in~\cite{Bonvin:2023jjq}.

\begin{figure}[h!]
\begin{center}
\includegraphics[clip,width=0.7\columnwidth]{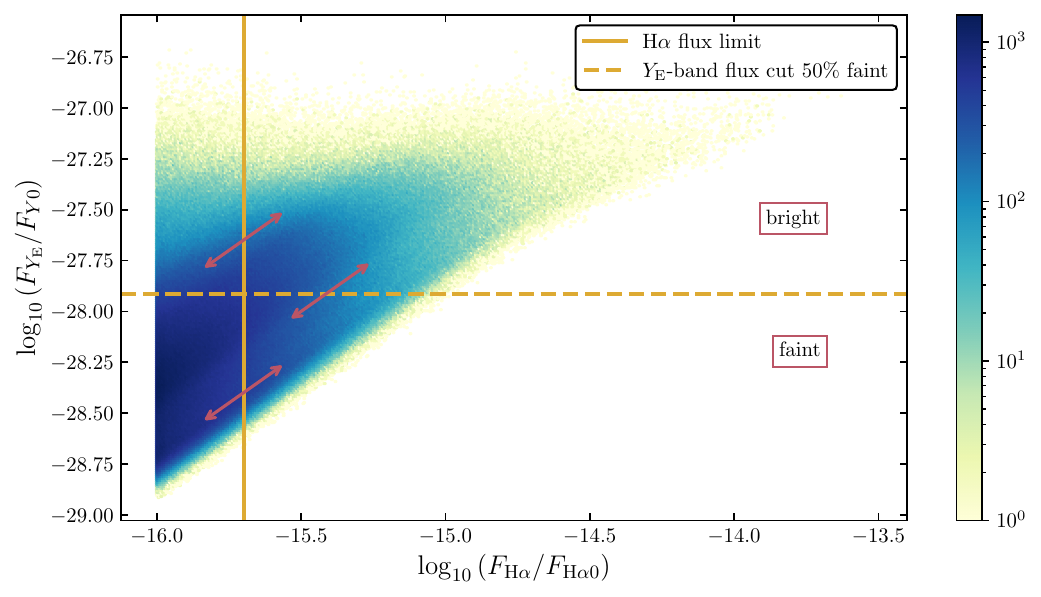}
\end{center}
\caption{Correlation between  $F_{\mathrm{H}\alpha}$ and $F_{Y}$ for the redshift bin centred at $\bar{z} = 1$ of the Flagship spectroscopic mock catalogue. A continuous line denotes the flux limit of the Euclid Wide Survey, while a dashed line highlights a \YE-band flux cut, splitting the sample in two equally-populated bright and faint populations. The reference fluxes are $F_{\mathrm{H}\alpha 0} = 1\,{\rm erg}\,{\rm cm}^{-2}\,{\rm s}^{-1}$ and $F_{Y0} \equiv 1\,{\rm erg}\,{\rm cm}^{-2}\,{\rm s}^{-1}\,{\rm Hz}^{-1}$. The colour bar highlights the density of objects (in arbitrary units). }
\label{fig:magn-flux}
\end{figure}

\Cref{fig:magn-flux} shows the physical intuition behind this computation. 
For each population of bright/faint objects, magnification can cause:
\begin{itemize} 
\item transfer of objects across the $F_{\mathrm{H}\alpha,\,\mathrm{lim}}$ threshold. Magnified galaxies at the edge of the threshold, with intrinsic luminosity below the survey limit, might be detected and classified as bright or faint, depending on their $Y$-band flux. This is represented in~\cref{fig:magn-flux} with the two arrows across the continuous line. The transfer between discarded objects and bright and faint populations is parameterised by the slope coefficients $s_{\mathrm{B}, \mathrm{H}\alpha}$ and $s_{\mathrm{F}, \mathrm{H}\alpha}$, respectively. These two coefficients are estimated from a version of the bright and faint catalogue with deeper coverage in H$\alpha$ flux, 
keeping the \YE-band flux selection fixed; 
\item transfer of objects across the $F_{\YE,\,\mathrm{lim}}$ cut between the faint and bright population. Objects that are intrinsically too faint to be classified as bright and are at the edge of the flux cut, if magnified, may be included in the bright population, and vice versa. This transfer is represented with the arrow across the dashed line and can be parameterised with the slope coefficient $s_{\mathrm{B}, \YE}$. This coefficient is estimated from the slope of the cumulative number of bright objects after applying the cut in the H$\alpha$ flux. 
\end{itemize}

The effective values of the local count slopes can also be estimated directly using the following procedure: 
\begin{enumerate}
    \item We magnify the fluxes by an arbitrary factor $\tilde{\mu}_{\rm L}$. For each object, we have
    \begin{equation}
     F_{\mathrm{H}\alpha} \rightarrow \tilde{F}_{\mathrm{H}\alpha} = \tilde{\mu}_{\rm L} \, F_{\mathrm{H}\alpha},  \qquad F_{\YE} \rightarrow \tilde{F}_{\YE}  = \tilde{\mu}_{\rm L}\, F_{\YE} \,.
    \end{equation}
    \item We repeat our selection, using the magnified fluxes. It is crucial at this point that the flux cut is exactly the same applied to the original fluxes. 
    \item Denoting with $N_{\mathrm{B}/\mathrm{F}}$ the number of objects in the bright/faint catalogue selected with the original fluxes, and with $\tilde{N}_{\mathrm{B}/\mathrm{F}}$ the number of objects selected with the magnified fluxes, we can estimate the effective local count slope as~\citep{Hildebrandt:2015kcb}
        \begin{equation}
        s^\mathrm{eff}_{\mathrm{B}/\mathrm{F}} = \frac{2}{5} \frac{\tilde{N}_{\mathrm{B}/\mathrm{F}}-N_{\mathrm{B}/\mathrm{F}}}{N_{\mathrm{B}/\mathrm{F}}} \frac{1}{\tilde{\mu}_{\rm L}-1}. 
    \end{equation}
    The result does not depend on the value of $\tilde{\mu}_{\rm L}$, for small values of $\tilde{\mu}_{\rm L}-1$. 
    Notice that $\tilde{\mu}_{\rm L}-1$ is the relative variation in flux that we have applied. Therefore, we are effectively computing 
    numerically the logarithmic derivative of the total number of objects in the bright/faint catalogue with respect to a flux variation, which is the definition of the local count slope.  
\end{enumerate}

\begin{figure}[h!]
\begin{center}
\includegraphics[clip,width=\columnwidth]{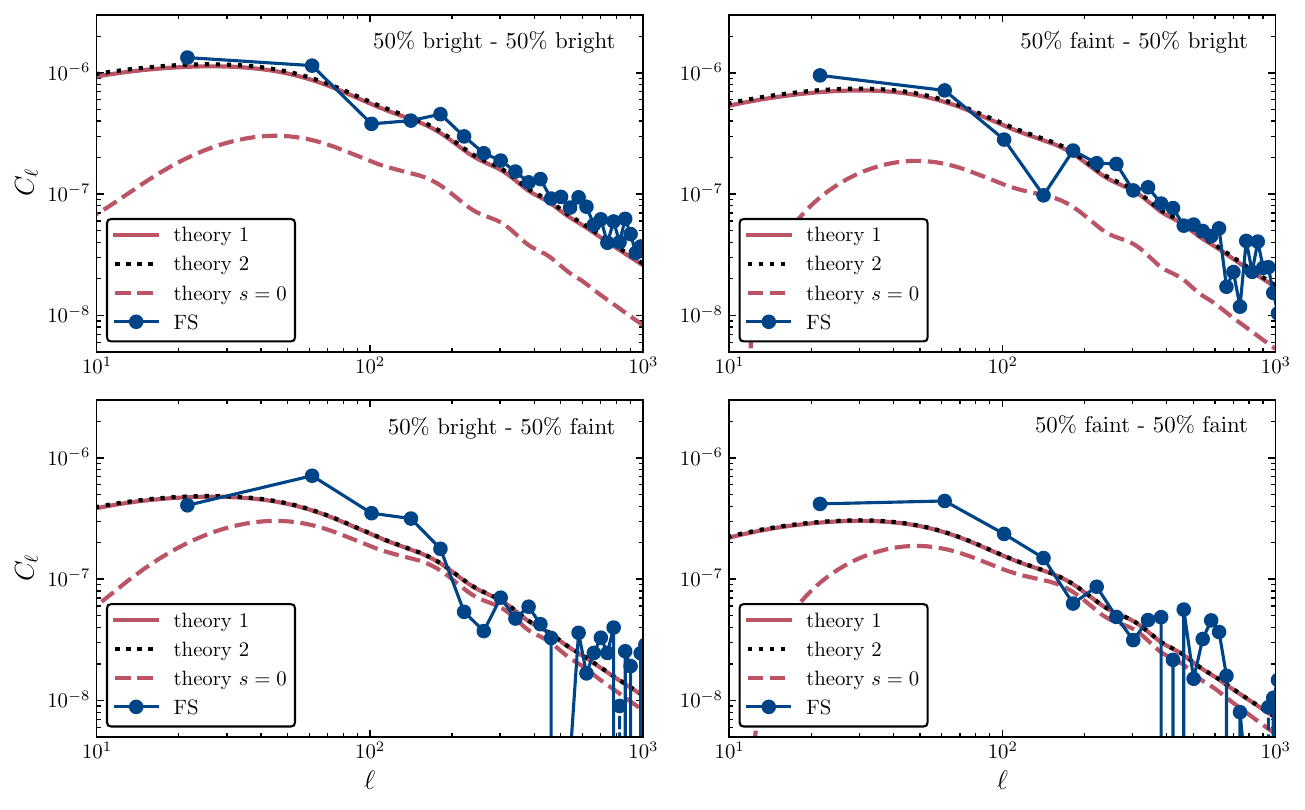}%
\end{center}
\caption{Angular power spectra for split 1 ($50\%$ of bright galaxies and $50\%$ of faint galaxies). Each panel represents a combination of maps: the top-left panel is the cross-correlation of the two number count 
maps of the bright sample, the top-right panel is the cross-correlation of the faint sample map at low redshift with the bright sample map at high redshift (both sensitive to the local count slope of the bright population),  the bottom-left panel is the cross-correlation of the bright sample map at low redshift with the faint sample map at high redshift, and the bottom-right panel is the cross-correlation of the two number count maps of the faint sample (both sensitive to the local count slope of the faint population). 
Data points represent simulation measurements, using the estimator defined in Eq.~\eqref{eq:cl-est}. 
Continuous red lines show the theory prediction using the analytical formula derived in this section for the local count slope, and black dotted lines are the predictions obtained from the values of the local count slope computed using the numerical differentiation method described in the text. A dashed line shows the absolute value of the theory model obtained with a local count slope equal to zero, that is, assuming no flux magnification (their true values are negative). }
\label{fig:test-magn-est-5050split}
\end{figure}

\begin{figure}[h!]
\begin{center}
\includegraphics[clip,width=\columnwidth]{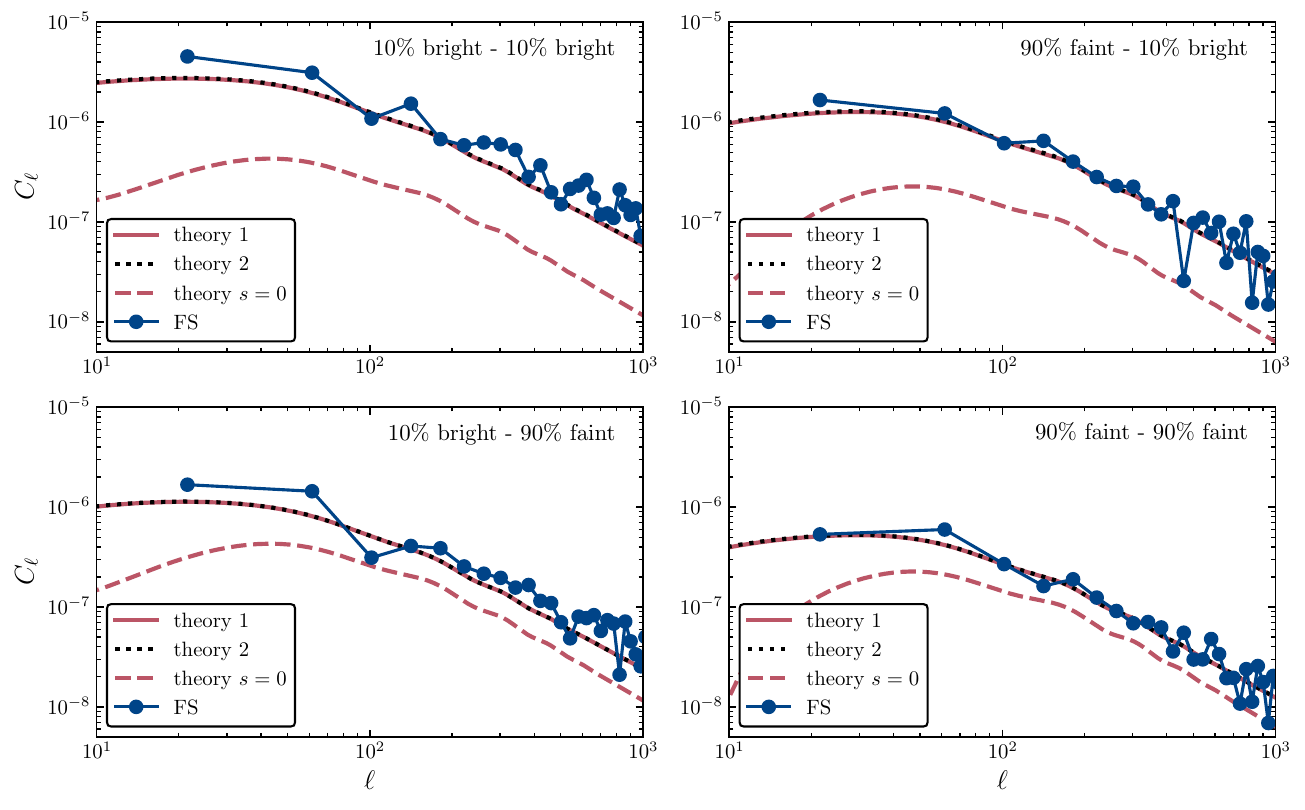}%
\end{center}
\caption{Same as in~\cref{fig:test-magn-est-5050split}, but for split 2 ($10\%$ of bright galaxies and $90\%$ of faint galaxies).}
\label{fig:test-magn-est-1090split}
\end{figure}

We compared the value of the local count slope using the two methods described above, finding agreement at the level of a few percent. 
We tested the consistency of both these measurements with the lensing magnification measured in the cross-correlation between redshift bins. 
As test cases, we considered the split with $50\%$ bright and $50\%$ faint objects (split 1), and the split with $10\%$ bright and $90\%$ faint objects (split 2). 
We constructed number count 
maps of the bright and faint populations. The binning in redshift was performed using the background redshift, so that
RSD effects are not present. We considered two types of maps: maps that are constructed from the comoving position of the
objects and selected based on their intrinsic flux (no flux magnificaton), and maps constructed from the magnified position of the objects and selected based on their magnified flux. 
The first case includes only contribution from the local density fluctuations of the galaxies, and the second case includes also lensing magnification. 

For each split, we estimated the angular cross spectrum by cross-correlating a map at low redshift ($\bar{z} = 1$) with a map at high redshift ($\bar{z} = 1.65$). 
In total, there are four possible configurations: (map of bright galaxies at $\bar{z} = 1$) $\times$  (map of bright galaxies at $\bar{z} = 1.65$),
(map of faint galaxies at $\bar{z} = 1$) $\times$ (map of bright galaxies at $\bar{z} = 1.65$), (map of bright galaxies at $\bar{z} = 1$) $\times$ (map of faint galaxies at $\bar{z} = 1.65$), and (map of faint galaxies at $\bar{z} = 1$) $\times$  (map of faint galaxies at $\bar{z} = 1.65$).
Following~\cite{Fosalba:2013mra}, we constructed an estimator for the magnified spectra $C^\mathrm{est}_\ell$, given by
\begin{equation}
C^\mathrm{est}_\ell = C^\mathrm{est, magn}_\ell - C^\mathrm{est, dens}_\ell, \label{eq:cl-est}
\end{equation}
where $C^\mathrm{est, magn}_\ell$ are the cross spectra estimated from the magnified maps, while $C^\mathrm{est, dens}_\ell$ are extracted from the density maps. 
The latter spectra are consistent with zero within statistical fluctuations, because there is no overlap between the two redshift bins. Therefore, this estimator removes the sample variance.

In~\cref{fig:test-magn-est-5050split,fig:test-magn-est-1090split} we compare the cross spectra estimated from the FS data with the two theory predictions for the local count slope that we have described above, for split 1 and split 2, respectively. 
Continuous red lines use values of the effective local count slope estimated from the analytical formula (theory 1), while black dotted lines represent the theory prediction with local count slope measured from the numerical derivative of the galaxy counts with respect to a variation of flux (theory 2).
The theory prediction has 
been predicted using the {\sc class} code~\citep{Blas:2011rf, CLASS, DiDio:2013bqa}, using the \texttt{HMCODE} to model nonlinearities~\citep{Mead:2016zqy}. 
For comparison, we also show in dashed red lines the absolute value of the theory prediction obtained setting the value of local count slope to zero (their true values are negative). While 
the latter prediction is completely off compared to the simulation measurements, the other two theory spectra are almost identical and agree very well with the data from FS. 
This cross-correlation, for all four combinations of maps,
is mostly dominated by the cross-correlation of the galaxy density at $\bar{z} = 1$ (sensitive to the galaxy bias) and lensing magnification at $\bar{z} = 1.65$ (sensitive to the local count slope). Therefore, it shows that the measurements of the local count slope are consistent with the magnification signal in the galaxy clustering angular statistics. 

\FloatBarrier 
\clearpage
\section{Properties of the galaxy populations}
\label{ap:table}

In~\Cref{tab:spec}, we report the properties of the galaxy populations we have used in our analysis.  

\begin{table}[h!]
\caption{\label{tab:spec} Specifics of the most representative galaxy samples. 
}
\begin{center}
\begin{tabular}{ c c c c c c c c c}
 \hline
split & \multicolumn{4}{c}{centrals} & \multicolumn{4}{c}{centrals + satellites} \\
 \hline
 & $N_\mathrm{gal}/10^6$ & $b$ & $s_\mathrm{eff}$ & $\overline{\Psi}^{\rm NFW}_0/c^2 \times 10^6$ & $N_\mathrm{gal}/10^6$ & $b$ & $s_\mathrm{eff}$ & $\overline{\Psi}^{\rm NFW}_0/c^2 \times 10^6$ \\
 \hline
 \hline
 \multicolumn{9}{c}{$\bar{z} = 1.0$} \\
 \hline
faint $90\%$&  $2.89$  & $1.42 \pm 0.02$ & $0.68 \pm 0.02$ & $-3.38$ & $3.25$ & $1.55 \pm 0.02$ & $0.72 \pm 0.03$ & $-3.76$ \\
bright $10\%$ & $0.321$ & $2.30 \pm 0.04$ & $1.44 \pm 0.06$ &  $-11.5$ & $0.361$ & $2.42 \pm 0.04$ & $1.45 \pm 0.05$ & $-11.7$ \\
\hline
faint $50\%$ &  $1.61$   & $1.25 \pm 0.01$ & $0.62 \pm 0.03$ & $-2.23$ & / &  / & / &  / \\
bright $50\%$ & $1.61$   & $1.75 \pm 0.03$ & $0.89 \pm 0.04$ & $-6.15$ & / & / & / & / \\
\hline
\hline
 \multicolumn{9}{c}{$\bar{z} = 1.2$} \\
 \hline
faint $90\%$&  $2.10$  &  $1.65 \pm 0.03$ & $0.76\pm 0.03$ & $-4.13$ & $2.48$ & $1.79 \pm 0.03$ & $0.80 \pm 0.03$ & $-4.55$ \\
bright $10\%$ &  $0.234$ &   $2.66 \pm 0.05$ & $1.54\pm 0.08$ &  $-12.8$ & $0.275$ & $2.78 \pm 0.05$ & $1.55 \pm 0.06$ & $-12.9$ \\
 \hline
faint $50\%$ &  $1.17$ &  $1.46 \pm 0.03$ & $0.68 \pm 0.03$ & $-2.86$ & / & / & / & / \\
bright $50\%$ & $1.17$   & $2.04 \pm 0.03$ & $0.99 \pm 0.02$ &  $-7.14$  & / & / & / & / \\
 \hline
\hline
 \multicolumn{9}{c}{$\bar{z} = 1.4$} \\
 \hline
faint  $90\%$&  $1.55$   & $2.0 \pm 0.03$ & $0.83\pm 0.03$ & $-4.66$ & $1.81$ & $2.13 \pm 0.03$ & $0.87 \pm 0.03$ & $-5.03$ \\
bright $10\%$ & $0.172$  & $3.19 \pm 0.07$ & $1.46\pm 0.4$ &  $-13.9 $ & $0.201$ & $3.24 \pm 0.06$ & $1.46 \pm 0.05$ & $-13.8$ \\
 \hline
faint $50\%$ &  $0.859$ &  $1.79 \pm 0.03$ & $0.71 \pm 0.04$ & $-3.30$ & / & / & / & / \\
bright $50\%$ & $0.859$  & $2.43 \pm 0.04$ & $1.05 \pm 0.03$ &  $-7.87$ & / & / & / & /  \\
\hline\hline
 \multicolumn{9}{c}{$\bar{z} = 1.65$} \\
 \hline
faint  $90\%$&  $1.53$  & $2.36 \pm 0.04$ & $0.89 \pm 0.02$ & $-6.16$ & $1.78$ & $2.59 \pm 0.04$ & $0.92 \pm 0.03$ & $-6.50$ \\
bright $10\%$ &  $0.170$ &   $3.61 \pm 0.08$ & $1.6 \pm 0.1$ &  $-15.4$ & $0.197$ & $3.70 \pm 0.08$ & $1.60 \pm 0.07$ & $-15.2$ \\
 \hline
faint $50\%$ &  $0.849$ & $2.11 \pm 0.04$ & $0.72 \pm 0.04$ & $-4.55$ & / & / & / & /  \\
bright $50\%$ & $0.849$ & $2.85 \pm 0.05$ & $1.18 \pm 0.05$ &  $-9.62$ & / & / & / & / \\
\hline
\hline
\end{tabular} 
\end{center}
\tablefoot{
Biases were measured in real space and neglect flux magnification. We report the dimensionless quantity $\overline{\Psi}^\mathrm{NFW}_0/c^2$, where $\overline{\Psi}^\mathrm{NFW}_0$ corresponds to the mean value of NFW potential at the galaxy positions for the selected objects.
We did not consider the split ($50\%$ faint, $50\%$ bright) for the case `centrals + satellites'. 
}
\end{table}

\end{appendix}

\end{document}